\documentclass[useAMS,usenatbib]{mn2e} %

\usepackage{amsmath, psfrag, color, epsfig, rotating, pstricks, pst-node}
\usepackage{dsfont}
\usepackage{txfonts}
\bibpunct{(}{)}{;}{a}{}{,}

\newcommand{\dd}{{\mathrm d}}
\newcommand{\J}{{\mathrm J}}

\renewcommand{\vec}[1]{\boldsymbol{#1}}
\newcommand{\vt}{\vartheta}

\newcommand{\Omegam}{\Omega_{\rm m}}

\newcommand{\Omegab}{\Omega_{\rm b}}

\newcommand{\rul}{\rule[-2.25mm]{0mm}{7mm}}

\newcommand{\RE}{{\cal R}_{\rm E}}
\newcommand{\RB}{{\cal R}_{\rm B}}
\newcommand{\RR}{\langle{\cal RR} \rangle_{\rm E}}
\newcommand{\rrx}{r}
\newcommand{\phm}{\phantom{-}}

\definecolor{darkgreen}{rgb}{0.1,0.5,0.1}

\title
{
  A new cosmic shear function: Optimised E-/B-mode decomposition on a finite
  interval
}
\author[Liping Fu and Martin Kilbinger]{Liping
  Fu$^{1,2}$\thanks{E-mail:    fu@oacn.inaf.it   } and Martin Kilbinger$^{3}$ \\
  $^{1}$INAF, Osservatorio Astronomico di Capodimonte, via Moiariello
  16, 80131 Napoli, Italy \\
  $^{2}$Shanghai Key Lab for Astrophysics, Shanghai Normal University,
  Shanghai 200234, P. R. China \\
  $^{3}$Institut d'Astrophysique de Paris, 98bis boulevard Arago,
  F-75014 Paris, France }

\begin{document}

\date{Accepted / Received}

\pagerange{\pageref{firstpage}--\pageref{lastpage}} \pubyear{2009}

\maketitle

\label{firstpage}

\begin{abstract}

  The decomposition of the cosmic shear field into E- and B-mode is an
  important diagnostic in weak gravitational lensing. However,
  commonly used techniques to perform this separation suffer from
  mode-mixing on very small or very large scales. We introduce a new
  E-/B-mode decomposition of the cosmic shear two-point correlation on
  a finite interval. This new statistic is optimised for cosmological
  applications, by maximising the signal-to-noise ratio (S/N) and a
  figure of merit (FoM) based on the Fisher matrix of the cosmological
  parameters $\Omegam$ and $\sigma_8$.

  We improve both S/N and FoM results substantially with respect to
  the recently introduced ring statistic, which also provides
  E-/B-mode separation on a finite angular range. The S/N (FoM) is
  larger by a factor of three (two) on angular scales between 1 and
  220 arc minutes. In addition, it yields better results than for the
  aperture-mass dispersion $\langle M_{\rm ap}^2 \rangle$, with
  improvements of 20\% (10\%) for S/N (FoM).  Our results depend on
  the survey parameters, most importantly on the covariance of the
  two-point shear correlation function. Although we assume parameters
  according to the CFHTLS-Wide survey, our method and optimisation
  scheme can be applied easily to any given survey settings and
  observing parameters. Arbitrary quantities, with respect to which
  the E-/B-mode filter is optimised, can be defined, therefore
  generalising the aim and context of the new shear statistic.
\end{abstract}


\section{Introduction}

Cosmic shear, the weak gravitational lensing effect induced on images
of distant galaxies by the large-scale structure in the Universe, has
become a standard tool for observational cosmology \citep[see][for
recent
reviews]{SaasFee,2008ARNPS..58...99H,2008PhR...462...67M}. Large
surveys have used cosmic shear to obtain measurements of the matter
density $\Omegam$ and the density fluctuation amplitude
$\sigma_8$. Recent constraints were obtained from ground-based surveys
such as CFHTLS\footnote{http://www.cfht.hawaii.edu/Science/CFHTLS}
\citep{JonBen07,FSHK08} and
GaBoDS\footnote{archive.eso.org/archive/adp/GaBoDS/DPS\_stacked\_images\_v1.0}
\citep{2007A&A...468..859H}. Space-based surveys like
COSMOS\footnote{cosmos.astro.caltech.edu}
\citep{2007ApJS..172..219L,2007ApJS..172..239M} and parallel ACS data
\citep{2007A&A...468..823S} advanced cosmic shear observations to very
small angular scales. Cosmic shear has contributed to constraining
dark energy \citep{JBBD06, KB09}. It is considered to be one of the
most promising method to shed light onto the origin of the recent
accelerated expansion of the Universe \citep{DETF,ESOFundWG}, and is a
major science driver for many future surveys like KIDS, Pan-STARRS,
DES, LSST, JDEM or Euclid.

One of the (few) diagnostics for cosmic shear analyses is the
decomposition of the shear field into its E- and B-mode. Most
commonly, the shear power spectrum or, equivalently, the shear
correlation function, is split into the gradient (E) and curl (B)
component \citep{2002ApJ...568...20C,
  2002A&A...389..729S}. Gravitational lensing produces, to first
order, a curl-free shear field and therefore, the presence of a B-mode
is an indication of residual systematics in the PSF correction and
shape measurement analysis.  To obtain competitive constraints on
cosmological parameters, in particular for dark-energy or
beyond-standard physics, galaxy shapes have to be determined to
sub-percentage precision. This requires excellent correction of PSF
effects arising from the atmosphere, telescope and camera
imperfections.

Apart from observational effects and measurement systematics, the
cosmic shear signal can be severely contaminated by intrinsic
correlations of galaxy orientation with other galaxies or their
surrounding dark matter structures \citep{2000MNRAS.319..649H}. This
occurs for galaxies at the same redshift e.g.~which reside in the same
dark halo (\emph{intrinsic alignment}). It also affects galaxies at
very different redshifts, where a background galaxy is lensed by
matter surrounding a foreground galaxy, therefore inducing a
\emph{shape-shear-correlation} between the two galaxies. An E-mode as
well as a B-mode arises from intrinsic alignment
\citep{2002ApJ...568...20C, 2002MNRAS.332..788M}.

Standard methods to separate the E- and B-mode power spectra (or
correlation functions) involve integrals up to arbitrary small or
large angular scales \citep{1998MNRAS.296..873S, 2002ApJ...568...20C,
  2002A&A...389..729S}.  However, shear correlations can be observed
only on a finite interval. Since the shear field is probed at galaxy
positions, the smallest usable scale is given by the confusion limit
for close galaxy pairs, which is typically several arc minutes for
ground-based surveys. The largest observed distance is for current
surveys several degrees. These limits make the \mbox{E-/B-}mode
separation imperfect, and a mixing of modes is induced at the 1-10\%
level \citep{KSE06}. To circumvent this shortcoming, a new
second-order function, the so-called ``ring statistic'', was
introduced which permits a clear E-/B-mode separation on a finite
interval \cite[][ hereafter SK07]{SK07}. In addition, the authors
developed conditions for general filter functions necessary for an
E-/B-mode decomposition for a finite angular range.

In this work, we present a method to find filter functions which
fulfill the SK07 conditions. We devise a scheme which provides an
optimised E-/B-mode decomposition on a finite interval.  The
optimisation is performed with regard to cosmological applications of
cosmic shear; the signal-to-noise ratio and a Fisher matrix figure or
merit are the quantities to be maximised.  This paper is organised as
follows: In Sect.~\ref{sec:method} we briefly review the results from
SK07 before we present our optimisation method. Results for the
signal-to-noise ratio and the figure of merit are shown in
Sect.~\ref{sec:results}. We conclude the paper with a summary
(Sect.~\ref{sec:summary}) and an outlook (Sect.~\ref{sec:outlook}).

\section{Method}
\label{sec:method}

\subsection{E-and B-mode Decomposition of the shear correlation
  function on a finite interval}
\label{sec:EB_finite}

We define the general second-order cosmic shear functions $\RE$ and $\RB$,
\begin{align}
  \RE & =  \frac 1 2 \int_0^\infty \dd \vt \, \vt \left[ T_+(\vt) \, \xi_+(\vt)
    + T_-(\vt) \, \xi_-(\vt) \right] \nonumber; \\
  \RB & =  \frac 1 2 \int_0^\infty \dd \vt \, \vt \left[ T_+(\vt) \, \xi_+(\vt)
    - T_-(\vt) \, \xi_-(\vt) \right],
  \label{EB}
\end{align}
as integrals over the shear two-point correlation functions $\xi_+$
and $\xi_-$ \citep[e.g.][]{1992ApJ...388..272K} with arbitrary filter
functions $T_+$ and $T_-$.  These expressions correspond to Eq.~(39)
from SK07, with an additional factor $1/2$ in our definition.  In
  terms of the E- and B-mode power spectrum, $P_{\rm E}$ and $P_{\rm
    B}$, respectively, the shear two-point correlation function are
  given as the following Hankel transforms
  \citep{2002A&A...389..729S}
\begin{align}
\xi_+(\vt)&=\int_0^\infty {\dd\ell\;\ell\over 2\pi}\,
{\rm J_0}(\ell\vt)[{P_{\rm E}(\ell)+P_{\rm B}(\ell)}]
\nonumber ;\\
\xi_-(\vt)&=\int_0^\infty {\dd\ell\;\ell\over 2\pi}\,
{\rm J_4}(\ell\vt)[{P_{\rm E}(\ell)-P_{\rm B}(\ell)}].
  \label{xi+-}
\end{align}
 with $\J_\nu$ being the $\nu^{\rm th}$-order Bessel function of the first kind.
Inserting Eq.~(\ref{xi+-}) into Eq.~(\ref{EB}) yields
\begin{align}
  \RE = \frac 1 2 \int_0^\infty \frac{\dd \ell \, \ell}{{2\pi}} \Bigg[
    & P_{\rm E}(\ell) \Big( {\cal W}_{\rm E}(\ell) + {\cal W}_{\rm
        B}(\ell) \Big) \nonumber \\
    +
    & P_{\rm B}(\ell) \Big( {\cal W}_{\rm E}(\ell) - {\cal W}_{\rm
        B}(\ell) \Big) \Bigg],
      \label{RE_power}
\end{align}
and an analogous expression for $\RB$. The Hankel transforms of $T_+$ and $T_-$ are defined as
\begin{align}
  {\cal W}_{\rm E, B}(\ell) = \int_0^\infty \dd \vt \, \vt \,
  T_{+,-}(\vt) \, \J_{0,4}(\vt \ell) .
\end{align}

To provide an E- and B-mode decomposition, in the sense that $\RE$
only depends on the E-mode of the shear field and $\RB$ only on its
B-mode, the two Hankel transform have to be identical, ${\cal W}_{\rm
  E} = {\cal W}_{\rm B}$. After some algebra, one finds that the
following equivalent relations between the filter functions $T_+$ and
$T_-$ must hold \citep{2002A&A...389..729S}
\begin{align}
  T_+(\vt) & = T_-(\vt) + 4 \int_{\vt}^\infty \frac{\dd \theta \,
    \theta}{\theta^2} \, T_-(\theta) \left[ 1 - 3
    \left(\frac{\vt}{\theta}\right)^2 \right];
  \label{Tp_from_Tm} \\
  T_-(\vt) & = T_+(\vt) + 4 \int_{0}^\vt \frac{\dd \theta \,
    \theta}{\vt^2} \, T_+(\theta) \left[ 1 - 3
    \left(\frac{\theta}{\vt}\right)^2 \right].
  \label{Tm_from_Tp}
\end{align}
Therefore, for an arbitrary function $T_+$, a corresponding filter
$T_-$ can be derived from $T_+$ to provide an E- and B-mode
decomposition, and vice versa. In the absence of a B-mode we have $\RB
= 0$, and $\RE$ can be obtained from $\xi_+$ or $\xi_-$ alone,
\begin{equation}
  \RE =  \int_0^\infty \dd \vt \, \vt \, T_+(\vt) \, \xi_+(\vt)
  = \int_0^\infty \dd \vt \, \vt \, T_-(\vt) \, \xi_-(\vt).
  \label{RE-noB}
\end{equation}

We further require $\RE$ and $\RB$ to depend on the shear correlation
given at angular scales $\vt$ in a finite interval, $0 < \vt_{\rm min}
\le \vt \le \vt_{\rm max} < \infty$. Thus, we demand $T_-$ to have
finite support $[\vartheta_{\rm min}; \vartheta_{\rm max}]$ and
$T_+(\vartheta)$ to vanish for $\vartheta<\vartheta_{\rm min}$,
Eq.~(\ref{Tp_from_Tm}) then implies the following integral constraints
on the filter function $T_-$ (SK07):
\begin{equation}
  \int_{\vt_{\rm min}}^{\vt_{\rm max}} \frac{\dd \vt}{\vt} T_-(\vt) =
  \int_{\vt_{\rm min}}^{\vt_{\rm max}} \frac{\dd \vt}{\vt^3} T_-(\vt)
  = 0.
  \label{c1c2minus}
\end{equation}
In addition, it follows that $T_+(\vartheta) = 0$ for
$\vartheta>\vartheta_{\rm max}$. Using the finite support of $T_+$ in
(Eq.~\ref{Tm_from_Tp}), we get integral constraints for $T_+$,
\begin{align}
  \int_{\vt_{\rm min}}^{\vt_{\rm max}} \dd \vt \, \vt T_+(\vt)
  =  \int_{\vt_{\rm min}}^{\vt_{\rm max}} \dd \vt \, \vt^3 T_+(\vt)
  = 0.
  \label{c1c2}
\end{align}
Then, $\RE$ and $\RB$ are functions of the two angular scales
$\vt_{\rm min}$ and $\vt_{\rm max}$. We will discuss the
scale-dependence in more detail in Sect.~\ref{sec:scale-dep}.

SK07 constructed a set of functions, $Z_+, Z_-$ in their notation,
which satisfy Eqs.~(\ref{c1c2minus}, \ref{c1c2}).  Those functions
were motivated from a geometrical ansatz, by considering two
concentric, non-overlapping rings. If the shear correlation is
calculated from galaxy pairs of which one galaxy lies in the inner
ring and the other galaxy in the outer ring, the E-/B-mode
decomposition on a finite interval is guaranteed by construction. The
form of the function $Z_+$ originated in a specific choice of the
weight profile over the two rings. The relation between $T_\pm$ and
$Z_\pm$ is
\begin{equation}
  T_\pm(\vt) = \vt^{-2} \, Z_\pm(\vt/\vt_{\rm max}).
\end{equation}
Note that in SK07 the analogous integrals to Eq.~(\ref{EB}) using
$Z_\pm$ are carried out over the integration variable $\vt/\vt_{\rm
  max}$ and extend from $\vt_{\rm min}/\vt_{\rm max}$ to 1.  The shear
second-order functions corresponding to Eq.~(\ref{EB}) are denoted as
$\langle {\cal RR} \rangle_{\rm E}$ and $\langle {\cal RR}
\rangle_{\rm B}$, respectively.

There are infinitely many functions which fulfill the above integral
constraints. Their choice can of course be detached from the
geometrical considerations of the ``ring statistic''. In this paper we
define a general filter function which we will optimise regarding some
specific criterion. This criterion will be related to the cosmological
information output from a cosmic shear survey. We will use two cases,
the signal-to-noise ratio and the Fisher matrix of cosmological
parameters. The results are presented in Sect.~\ref{sec:results}.

\subsection{Parametrisation of the filter function}


For the optimisation problem, we focus on $T_+$ since $T_-$ can be
derived from $T_+$ (Eq.~\ref{Tm_from_Tp}). First, we remap $T_+$ to
the interval $[-1;+1]$ by defining
\begin{align}
  \tilde T_+(x) = & \, T_+(Ax + B) = T_+(\vt) \quad \mbox{for} \quad x
  \in [-1; 1]; \nonumber \\
  A = & \, (\vt_{\rm max} - \vt_{\rm min})/2; \;\;\; B = (\vt_{\rm max} +
  \vt_{\rm min})/2.
  \label{theta2x}
\end{align}
With that the two integral constraints (Eq.~\ref{c1c2}) become
\begin{align}
  \int_{-1}^{+1} \dd x \, (x + R) \, \tilde T_+(x)
  = \int_{-1}^{+1} \dd x \, (x + R)^3 \, \tilde T_+(x)
  = 0,
  \label{c1c2x}
\end{align}
where we have defined the ratio $R$ as
\begin{equation}
  R = \frac B A = \frac{1 + \eta}{1 - \eta}; \;\;\;\; \eta = \frac{\vt_{\rm
      min}}{\vt_{\rm max}}.
\end{equation}


Next, we decompose $\tilde T_+$ into a finite sum of orthogonal
polynomials:
\begin{equation}
  \tilde T_+(x) = \sum_{n=0}^{N-1} a_n \, C_n(x).
  \label{decomp}
\end{equation}
This representation allows us to find an optimal filter function by
varying the coefficients $a_n$.  Sect.~\ref{sec:optimisation}.  The
polynomials $C_n$ can be chosen freely; we use Chebyshev polynomials
of the second kind,
\begin{equation}
  {\rm U}_n(x) = \frac{\sin [(n+1) \arccos
      x]}{\sin(\arccos x)} .
  \label{cheby2}
\end{equation}
The optimisation process is then performed by varying the coefficients
$a_n$; this is described in detail in Sect.~\ref{sec:optimisation}.

Apart from the integral constraints (Eq.~\ref{c1c2}), {one could}
require $T_+$ to be zero at the interval boundaries,
\begin{align}
  \mbox{Continuity}: \quad & T_+(\vt_{\rm max}) = T_+(\vt_{\rm min}) =
  0.
\label{Continuity}
\end{align}
Additional constraints could be added, for example differentiability
at the boundaries. We will discuss their effects on the results in
Sect.~\ref{sect:constraints}.

\subsection{Satisfying the constraints}
\label{sec:constraints}

In general, the function $T_+$, or equivalently $\tilde T_+$, is
constrained by $K \ge 2$ equations in the form $F_m[\tilde T_+] = 0,\;
m=0 \ldots K-1$. If the functionals $F_m$ are linear in $\tilde T_+$,
applying Eq.~(\ref{decomp}) leads to
\begin{equation}
  \sum_{n=0}^{N-1} f_{mn} \, a_n = 0; \;\;\;\; f_{mn} := F_m[C_n].
  \label{faeq0}
\end{equation}
The matrix element $f_{mn}$ is the $m^{\rm th}$ constraint applied to
the orthogonal polynomial of order $n$. For example, taking the first
constraint in Eq.~(\ref{c1c2x}) we get
\begin{equation}
  f_{0n} = \int_{-1}^{+1} \dd x \, (x + R) \, C_n(x),
\end{equation}
which can be integrated analytically. The other matrix elements
$f_{mn}$ are obtained analogously.

$K$ constraints fix $K$ coefficients of the decomposition
(Eq.~\ref{decomp}), the remaining $N-K$ coefficients can be chosen
arbitrarily. We use the highest $N-K$ coefficients as free parameters
($n=K \ldots N-1$) and fix the first $K$ coefficients ($n=0 \ldots
K-1$) as follows. Define
\begin{equation}
  s_m = - \sum_{n=K}^{N-1} f_{mn} \, a_n; \;\;\;\; m = 0 \ldots K-1,
  \label{s_m}
\end{equation}
 Eq.~(\ref{faeq0}) can then be written as
\begin{equation}
  \sum_{n=0}^{K-1} f_{mn} \, a_n = s_m; \;\;\;\; m = 0 \ldots K-1.
  \label{eqK}
\end{equation}
This $(K\times K)$-matrix equation is solved for the first $K$
coefficients $a_n \, (n=0\ldots K-1)$ by inverting the square
(sub-)matrix $(f_{mn})_{m,n<K}$ on the left-hand side of the above
equation.  If the constraints are chosen such that they are linearly
independent (which is this case), this matrix equation has a unique,
non-trivial solution.

In addition to the $K$ constraints, we impose an integral
normalisation of the filter function given by the $L^2$-norm
\begin{equation}
  ||\tilde T_+||^2_2 = \int_{-1}^1 \dd x \, w(x) \tilde T_+^2(x) = 1,
\label {normalisation}
\end{equation}
where $w$ is the corresponding weight of the polynomial family, $w(x)
= (1-x^2)^{1/2}$ in the case of second-kind Chebyshev polynomials.
This normalisation does not affect the constraints which are
independent of a multiplication of all $a_n$ with a common
factor. Note also that the quantities which we will optimise in
Sect.~\ref{sec:results} do not depend on the normalisation.

\subsection{Calculation of $T_-$ from $T_+$}
\label{sec:Tminus}

To obtain the function $\tilde T_- := T_-(A x + B)$ we transform
Eq.~(\ref{Tm_from_Tp}) to
\begin{align}
  \tilde T_-(x) = & \, \tilde T_+(x) + 4 \int_{-1}^x \dd x^\prime \,
  \tilde T_+(x^\prime) \frac{x^\prime + R}{(x + R)^2} \,  \left[ 1 - 3
    \left(\frac{x^\prime + R}{x + R}\right)^2 \right]; \nonumber \\
   & x = -1 \ldots 1,
  \label{Tm_tilde}
\end{align}
which can be written as
\begin{align}
  \tilde T_-(x) = & \sum_{n=0}^{N-1} a_n \left[ C_n(x) + \alpha_n(x)
  \right] ; \\
  \alpha_n(x) = & \, 4 \int_{-1}^x \dd x^\prime \,
  C_n(x^\prime) \frac{x^\prime + R}{(x + R)^2} \, \left[ 1 - 3
    \left(\frac{x^\prime + R}{x + R}\right)^2 \right],
  \label{alpha}
\end{align}
inserting the decomposition (Eq.~\ref{decomp}). We define
\begin{equation}
  F_n^{(\nu)}(x) = \int_{-1}^x \dd x^\prime \, (x^\prime)^{\nu}
  C_n(x^\prime); \;\;\;\; \nu = 0,1,2,3,
  \label{F_n_nu}
\end{equation}
and write Eq.~(\ref{alpha}) as
\begin{align}
  \alpha_n(x) = & \frac{4}{3} \rrx \left[ R(\rrx-R) F_n^{(0)} + (1 - 2 \rrx
    R - \rrx R^2) F_n^{(1)} \right. \nonumber \\
    &
    \left. - \rrx R (R+2) F_n^{(2)} - \rrx F_n^{(3)} \right];
  \nonumber \\
  r = & \frac{3}{(x + R)^2},
  \label{alpha2}
\end{align}
where we dropped the argument $x$ from $F_n^{(\nu)}$ and $r$.  The
integral (Eq.~\ref{F_n_nu}) for $\nu=0$ and $C_n = {\rm U}_n$ is
\begin{align}
  F_n^{(0)}(x) = \left\{ \begin{array}{ll}
    \frac{\displaystyle (-1)^n + x \, {\rm T}_n(x) - (1-x^2)
      {\rm U}_{n-1}(x)}{\displaystyle n+1}
    & (n \ne -1) \\[1.0em]
    0 & (n = -1)
    \end{array} \right.,
  \label{Fn0_U}
\end{align}
where ${\rm T}_n$ is $n^{\rm th}$-order Chebyshev polynomial of the first kind,
\begin{equation}
  {\rm T}_n(x) = \cos(n \arccos x).
  \label{cheby1}
\end{equation}

By using the recurrence relation of the Chebyshev polynomials, we
obtain the other functions,
\begin{align}
  F_n^{(1)} = & \frac 1 2 \left[ F_{n+1}^{(0)} + F_{n-1}^{(0)}
    \right]; \nonumber \\
  F_n^{(2)} = & \frac 1 4 \left[ F_{n+2}^{(0)} + 2
    F_{n}^{(0)} + F_{n-2}^{(0)}
    \right]; \\
  F_n^{(3)} = & \frac 1 8 \left[ F_{n+3}^{(0)} + 3
    F_{n+1}^{(0)} + 3
    F_{n-1}^{(0)} +  F_{n-3}^{(0)}
    \right]. \nonumber
\end{align}
Note that Eq.~(\ref{Fn0_U}) is valid for integer $n$, since the
expressions for the orthogonal polynomials are well-defined for
negative $n$.

With that, it can be readily checked whether the coefficients $a_n$
obtained with the method described in Sect.~\ref{sec:constraints} and
the resulting filter functions $\tilde T_\pm$ indeed provide an
E-/B-mode decomposition, by verifying that the B-mode $\RB$
(Eq.~\ref{EB}) is zero. In Sect.~\ref{sec:Bmode} we discuss numerical
issues when calculating the B-mode.

\subsection{Relation to the lensing power spectrum}
\label{sec:powerspectrum}

Eq.~(\ref{RE_power}) shows the relation of $\RE$ to the power
  spectrum. Assuming a pure E-mode, this equation reads
\begin{equation}
\RE = \frac 1 {2\pi} \int_0^\infty \dd \ell \, \ell P_{\rm E}(\ell)
\, {\cal W}_{\rm E}(\ell) ,
\label{RE_P}
\end{equation}
where the Fourier-space filter function $\cal W_{\rm E}$ can be written as
\begin{align}
  {\cal W}_{\rm E}(\ell, \vt_{\rm min}, \vt_{\rm max}) & = \sum_{n=0}^{N-1}
  a_n {\cal W}_n(\ell, \vt_{\rm min}, \vt_{\rm max});
  \label{W} \\
  {\cal W}_n(\ell, \vt_{\rm min}, \vt_{\rm max}) & = \int_{\vt_{\rm min}}^{\vt_{\rm max}} \dd \vt \, \vt
  \, C_n[(\vt-B)/A] \, \rm J_0(\vt \ell),
  \label{Wn}
\end{align}
with $A$ and $B$ given in Eq.~(\ref{theta2x}).
Analogously, $\RB$ can be written in terms of the B-mode power
spectrum $P_{\rm B}$
\begin{equation}
\RB = \frac 1 {2\pi} \int_0^\infty \dd \ell \, \ell P_{\rm B}(\ell)
\, {\cal W}_{\rm B}(\ell, \vt_{\rm min}, \vt_{\rm max}),
\end{equation}
 with ${\cal W}_{\rm E} = {\cal W}_{\rm B}$, see Sect.~\ref{sec:EB_finite}.

Unfortunately there is no simple analytical expression of
Eq.~(\ref{Wn}), which would be desirable to calculate Eq.~(\ref{RE_P})
using a fast Hankel transform \citep[FHT,][]{2000MNRAS.312..257H}. For
the moment, the most efficient method to calculate $\RE$ from a model
power spectrum is to obtain $\xi_+$ from $P_{\rm E}$ by FHT and to
integrate it via Eq.~(\ref{RE-noB}) which is fast since in our case
$T_+$ is a rather low-order polynomial, as we will see later.

\subsection{Angular-scale-dependence}
\label{sec:scale-dep}

The two constraints (Eq.~\ref{c1c2}) depend on the ratio of angular
scales $\eta = \vt_{\rm min}/\vt_{\rm max}$. A filter function $\tilde
T_+$ with given coefficients $a_n$ satisfying those constraints does
not in general fulfill the same constraints for a different
$\eta$. This therefore causes the inconvenience of having a different
filter function for each angular scale.

It should be noted that although formally $\RE$ is a function of two
angular scales, the information about cosmology and large-scale
structure is captured by a single parameter, let us call it
$\lambda$. This is because the large-scale matter power spectrum, of
which $\RE$ is a logarithmic convolution
(Sect.~\ref{sec:powerspectrum}), only depends on a single scalar. We
therefore expect a large covariance between many pairs $(\vt_{\rm
  min}, \vt_{\rm max})$. Although there are infinitely many mappings
$(\vt_{\rm min}, \vt_{\rm max}) \rightarrow \lambda$, two ways to
handle the scale-dependence seem to be suitable: (1) Leaving one scale
constant, and varying the other scale, e.g.~$\vt_{\rm min} =
\mbox{const.}, \lambda = \vt_{\rm max}$. (2) Leaving the ratio of both
scales constant, $\lambda = \eta$. We will pursue both ways in this
paper.

\subsubsection{Fixed minimum scale, $\vt_{\rm min} =  {\rm const}$}

The first mapping introduced above offers a rather efficient sampling
of the shear field. $\vt_{\rm min}$ can be fixed to the smallest
observable distance $\vt_{\rm min, 0}$ for which shear correlation
data are measured. This is given by the smallest separation for which
galaxies do not blend, to allow for reliably measured shapes. It
provides the largest range of angular scales accessible for a given
$\vt_{\rm max}$: The upper limit $\vt_{\rm max}$ can be varied between
$\vt_{\rm min}$ and the maximum observed scale given by the data.

\subsubsection{Fixed ratio, $\eta = \vt_{\rm  min}/\vt_{\rm max} ={\rm
    const}$}

The second mapping has the advantage that a single filter function can
satisfy the constraints (Eq.~\ref{c1c2}) for all scales. This makes it
more convenient to combine different scales, e.g.~to obtain the Fisher
matrix, and might result in a universal optimal filter function. The
efficiency with respect to keeping $\vt_{\rm min}$ constant is however
reduced: A large ratio $\eta$, for which $\vt_{\rm min}$ and $\vt_{\rm
  max}$ are close, samples only a small angular interval, resulting in
a small signal-to-noise. A small $\eta$ on the other hand means that
we can not go to very small scales with $\vt_{\rm max}$: Because of
the minimum observable galaxy separation $\vt_{\rm min, 0}$, the
smallest $\vt_{\rm max}$ is $\mbox{min}(\vt_{\rm max}) = \vt_{\rm min,
  0}/\eta$.

In both cases, we will use $\vt_{\rm max}$ as the argument of $\RE$
and denote it with the symbol $\Psi$, as in SK07.

In Appendix~\ref{sec:app}, we introduce a simple generalisation of
this scheme to obtain an optimised function $\tilde T_+$ which
fulfills the integral constraints (Eq.~\ref{c1c2}) for all pairs
$(\vt_{\rm min}, \vt_{\rm max})$. However, the corresponding
signal-to-noise ratio is significantly lower than with the method used
here, which was presented in Sect.~\ref{sec:constraints}. We therefore
do not consider this option further.

A remark about the analogy to the aperture-mass dispersion $\langle
M_{\rm ap}^2 \rangle$ is appropriate here. $\langle M_{\rm ap}^2
\rangle$ is obtained from the correlation function in a similar way as
$\RE$ in Eq.~(\ref{EB}), with integration range between zero and twice
the aperture radius. Its filter function depends on the two scales
$\vt$ (the integration variable) and $\theta$ (the aperture
radius). For different $\theta$ one could define a different filter
function. For convenience however, widely-used filters are functions
of the ratio $\vt/\theta$, and therefore one functional form provides
an E-/B-mode separation for all radii simultaneously.

\begin{figure*}
  \resizebox{\hsize}{!}{
    \includegraphics[angle=270,bb=50 100 554 760]{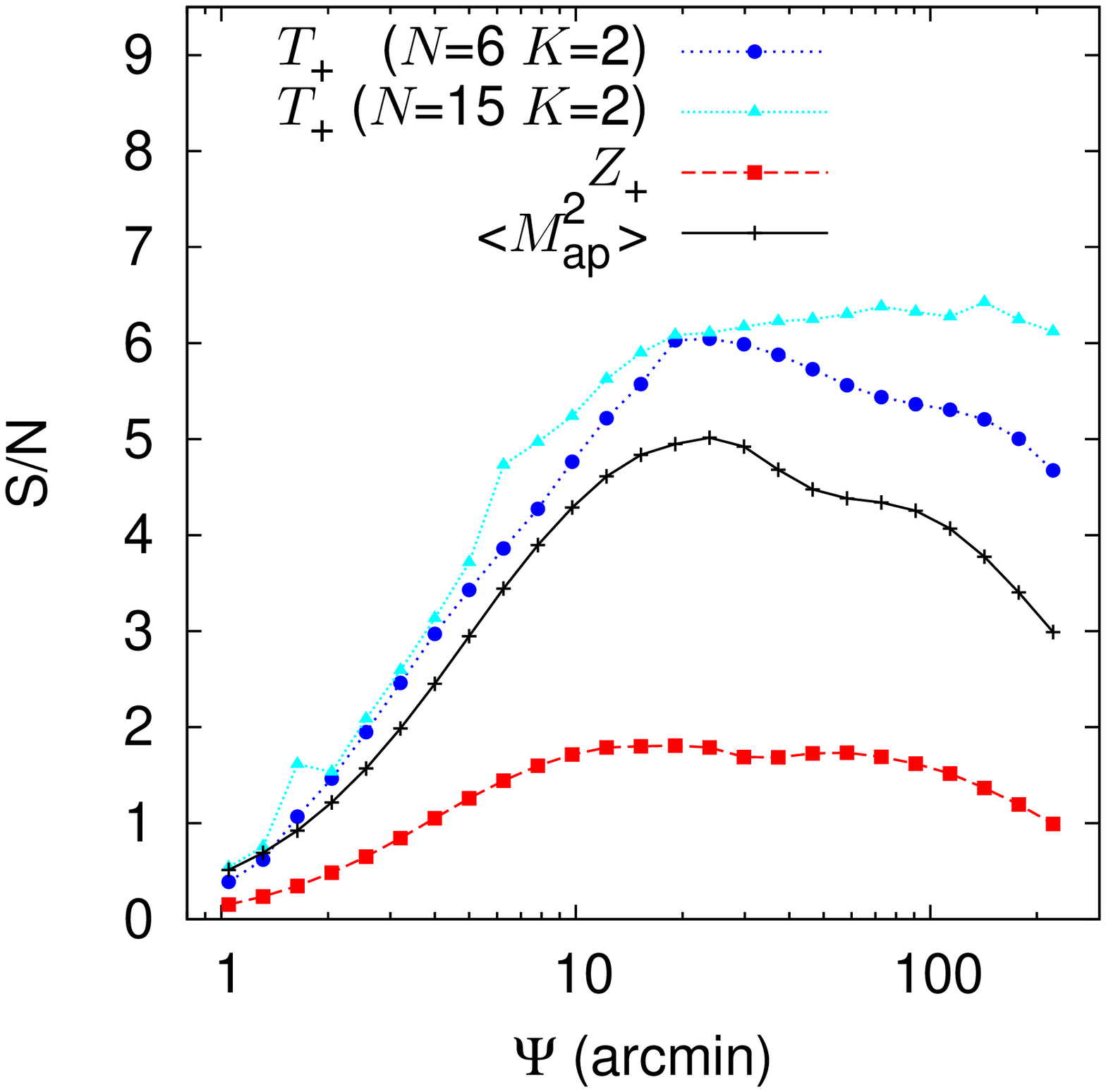}
    \includegraphics[angle=270,bb=50 115 554 770]{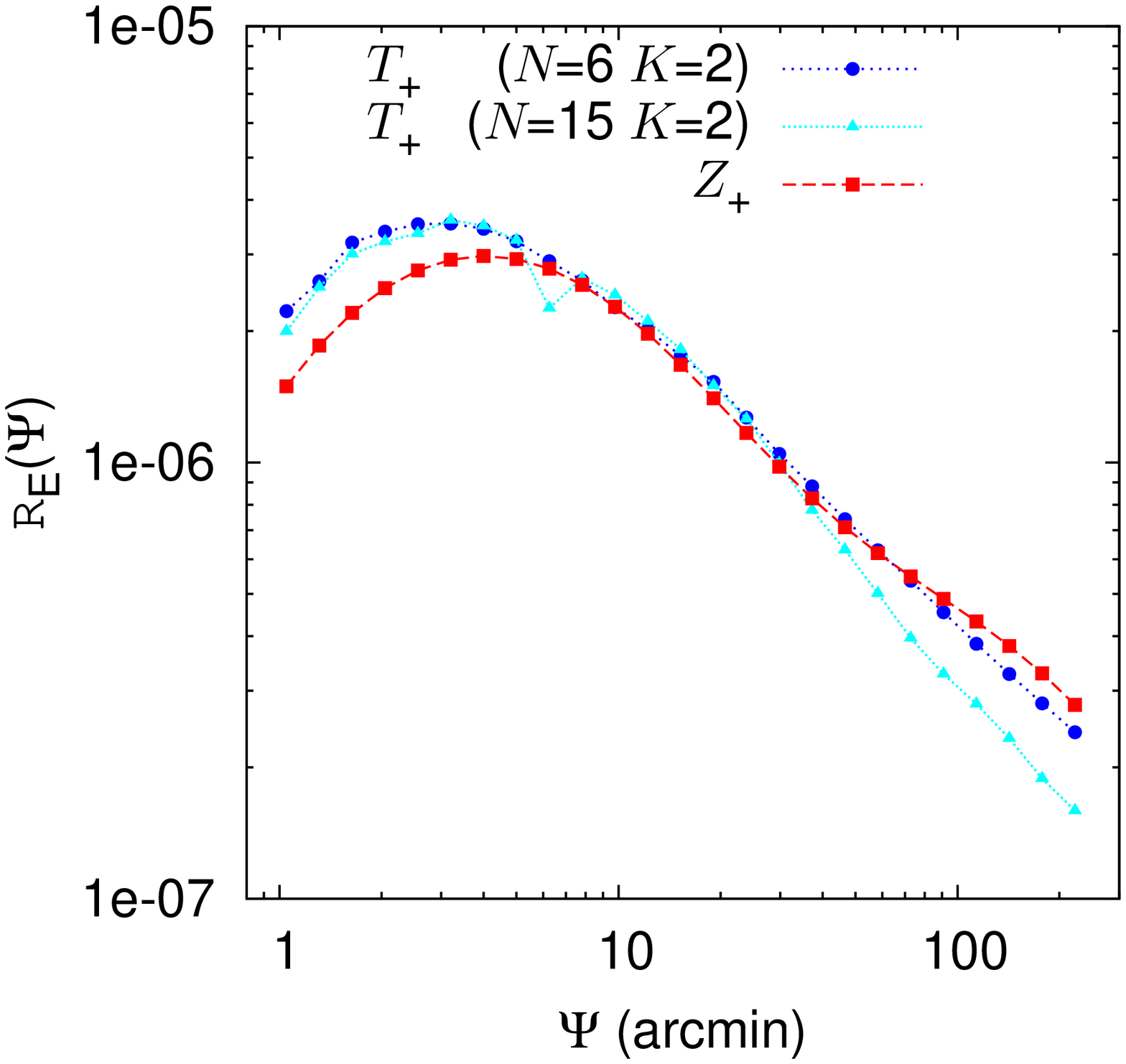}
  }
  \caption{\emph{Left panel}: Signal-to-noise ratio for fixed
    $\vt_{\rm min}=0.2\arcmin$, as functions of the maximum scale
    $\Psi$. The blue curve with circles and the cyan with triangles
    correspond to polynomial order $N=6$ and $N=15$, respectively. The
    red curve with squares shows the filter function from SK07. The
    black curve with crosses corresponds to the aperture-mass $\langle
    M_{\rm ap}^2 \rangle$ with $\Psi$ being equal to the aperture
    diameter. \emph{Right panel}: The comparison of $\RE$ obtained
    from the optimised function $T_+$ with polynomial orders $N=6$ and
    15. }

  \label{fig:SN_fixed_thetamin}
\end{figure*}

\subsection{Optimisation}
\label{sec:optimisation}

The maximisation of a quantity $Q$, to be defined in the next section
as signal-to-noise ratio and Fisher matrix figure of merit, is done as
follows. For a given polynomial order $N$ number of constraints $K \ge
2$, we perform the maximum search of $Q$ using the conjugate-gradient
method \cite{nr} in the space of free coefficients $a_K, \ldots,
a_{N-1}$. At each step, we calculate $s_m$ (Eq.~\ref{s_m}) and invert
Eq.~(\ref{eqK}) to get the first $K$ coefficients $a_0, \ldots
a_{K-1}$; from there we compute $\tilde T_+$ (Eq.~\ref{decomp}) and
$\RE$ (Eq.~\ref{RE-noB}).

We limit each of the coefficients $a_K, \ldots, a_{N-1}$ to the box
$[-10; 10]$.  In some cases the maximum-search fails, in particular if
the polynomial order is high. The algorithm might run into a local
maximum or hit the parameter boundary. To reduce the failure rate, we
proceed as follows.

If a maximum is found close to the parameter boundary we discard it
and redo the maximisation with larger box size for a different
starting point. To initialise the maximisation for the first of a
range of angular scales, we draw a number of random points, on the
order 100, and start the maximisation with the point providing the
largest $Q$.

For subsequent angular scales, we use the information about the
previous maximum to start the next optimisation. If the ratio $\eta =
\vt_{\rm min}/\vt_{\rm max}$ is kept constant when increasing
$\vt_{\rm max} = \Psi$, we use the previous maximum coefficients $a_n$
as new starting value for the maximisation process. This renders the
search for the maximum point more efficient and more stable, since for
small changes in $\Psi$ the maximum will be close in $a$-space.

For constant minimum scale $\vt_{\rm min}$ we devise a different
strategy. It can be shown that in this case, $Q$ is monotonously
increasing with $\Psi$: Let $T_+^{(i)}$ be the function for scale
$\Psi_i$ which maximises $Q(\Psi_i)$. For the subsequent scale
$\Psi_{i+1}>\Psi_i$, define the function $\bar T_+(\Psi) =
T_+^{(i)}(\Psi)$ for $\Psi<\Psi_i$, and $\bar T_+(\Psi) = 0$
otherwise. By construction, the resulting $Q$ for scale $\Psi_{i+1}$
is the same as for $\Psi_i$, $Q(\Psi_{i+1}) = Q(\Psi_i)$.  Therefore,
we choose as new starting point for scale $\Psi_{i+1}$ the
coefficients resulting from the decomposition of the function $\bar
T_+(\Psi)$. This assures the subsequent maximum $Q(\Psi_{i+1})$ found
by the search algorithm to be larger or equal to the previous one, at
least in the limit of large $N$. In practice however, due to the
finite order $N$, the orthogonal polynomials are not a good
representation of $\bar T_+$ for $\Psi>\Psi_i$ where it is zero, and
the resulting $Q$ can actually decrease with increasing scale.

We summarise the the combinations of $Q$, the angular dependencies and
the choice of the starting point for subsequent scales in Table
\ref{tab:optimise}.

\begin{table}
  \caption{Overview of quantities kept fixed when varying the
      scales $\Psi$, and the corresponding method to determine the
      starting point for subsequent scales. }

  \begin{tabular}{lll}
    \hline\hline $Q$ & Fixed quantity & Starting point for subsequent
    scales \\ \hline
    S/N & $\vt_{\rm min} = \mbox{const.}$ & Previous maximum function, $\bar T_+$ \\
    S/N & $\eta = \mbox{const.}$ & Previous maximum coefficients $a_n$
     \\
    FoM &  $\eta = \mbox{const.}$  & Previous maximum coefficients $a_n$   \\ \hline\hline
  \end{tabular}
 \label{tab:optimise}
\end{table}

\begin{figure*}
  \begin{center}
    \resizebox{\hsize}{!}{
      \includegraphics[angle=270,bb=50 50 540 340]{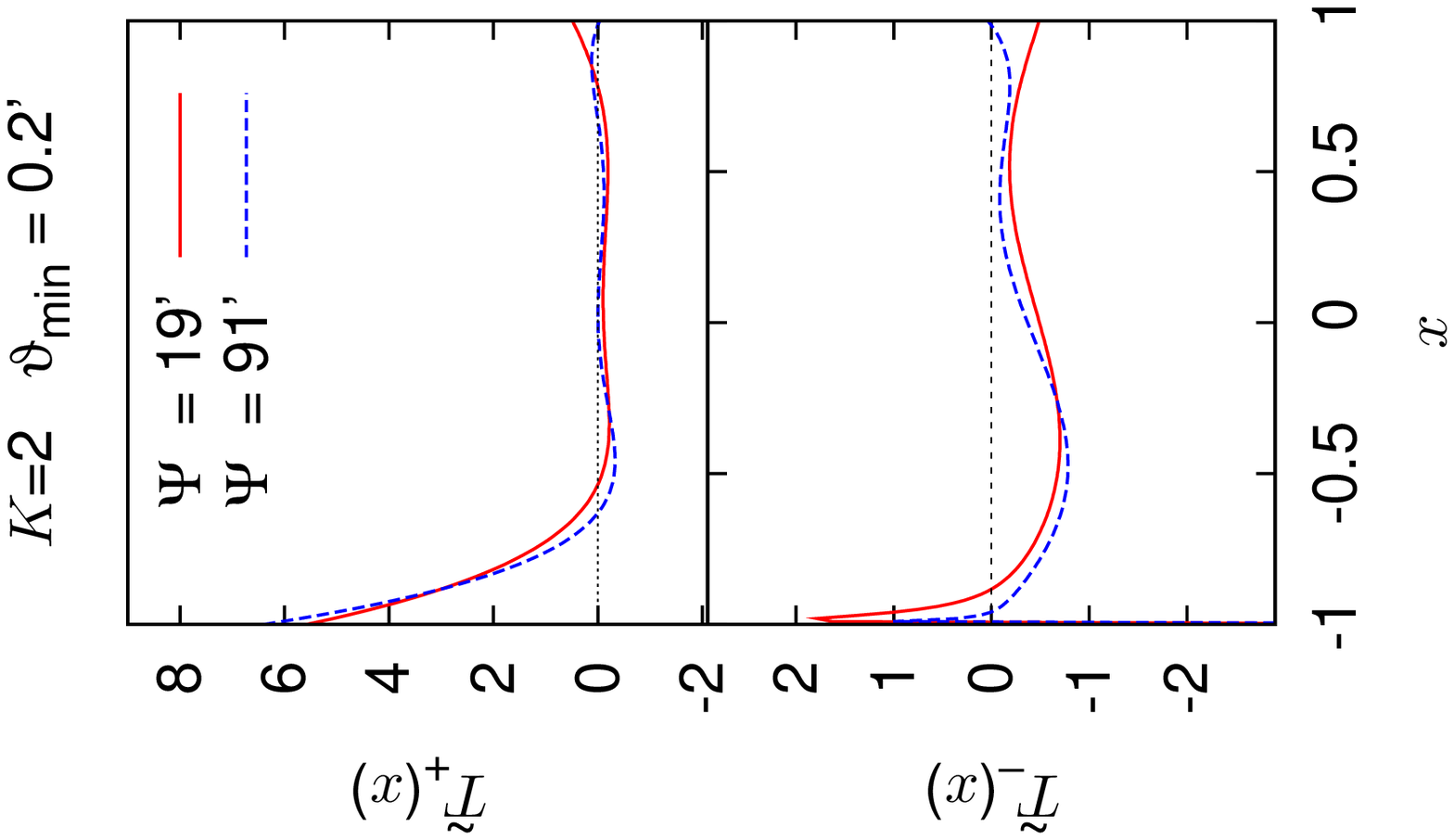}
      \includegraphics[angle=270,bb=50 50 540 340]{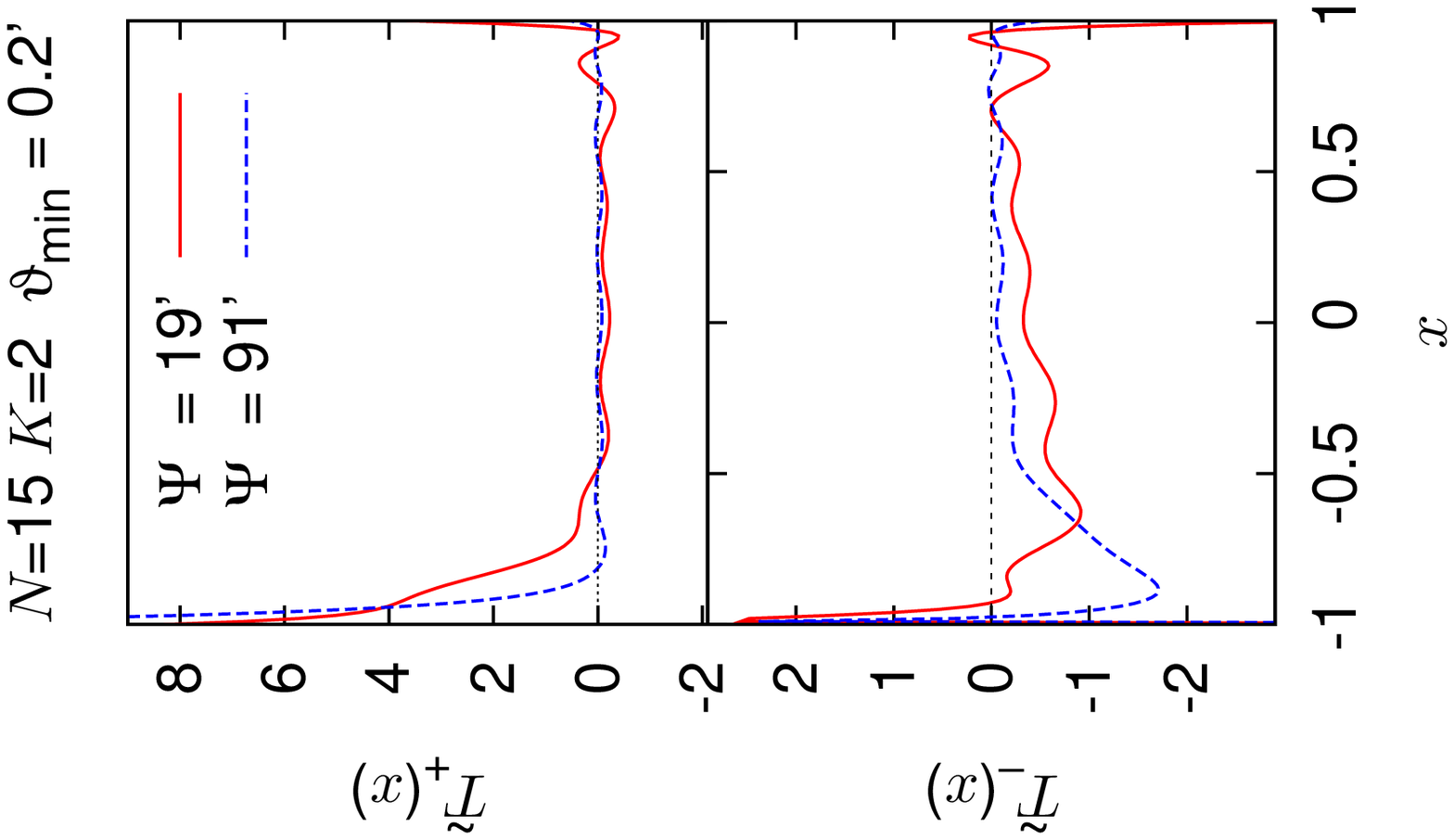}
      \includegraphics[angle=270,bb=50 50 540 340]{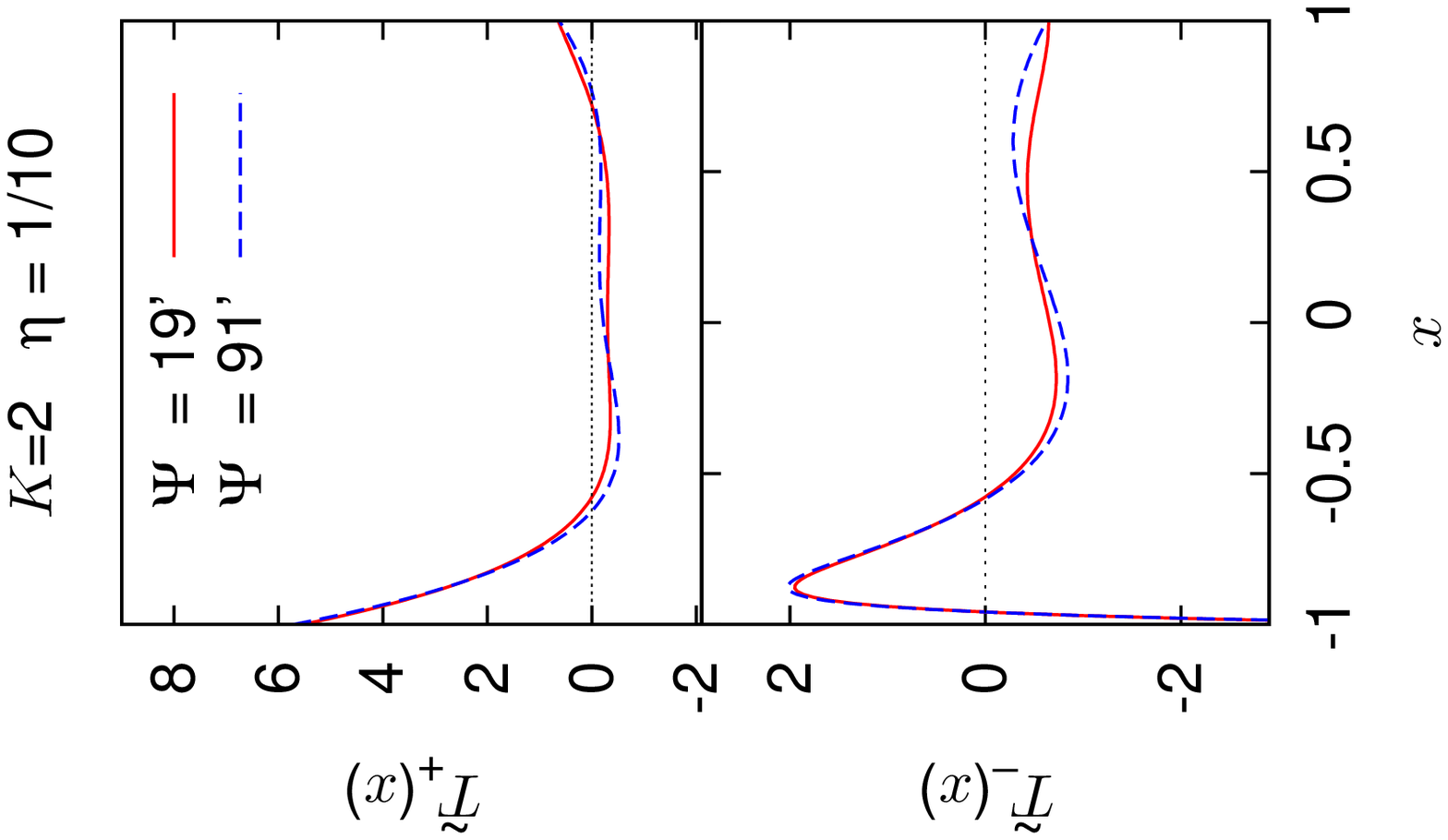}
      \includegraphics[angle=270,bb=50 50 540 340]{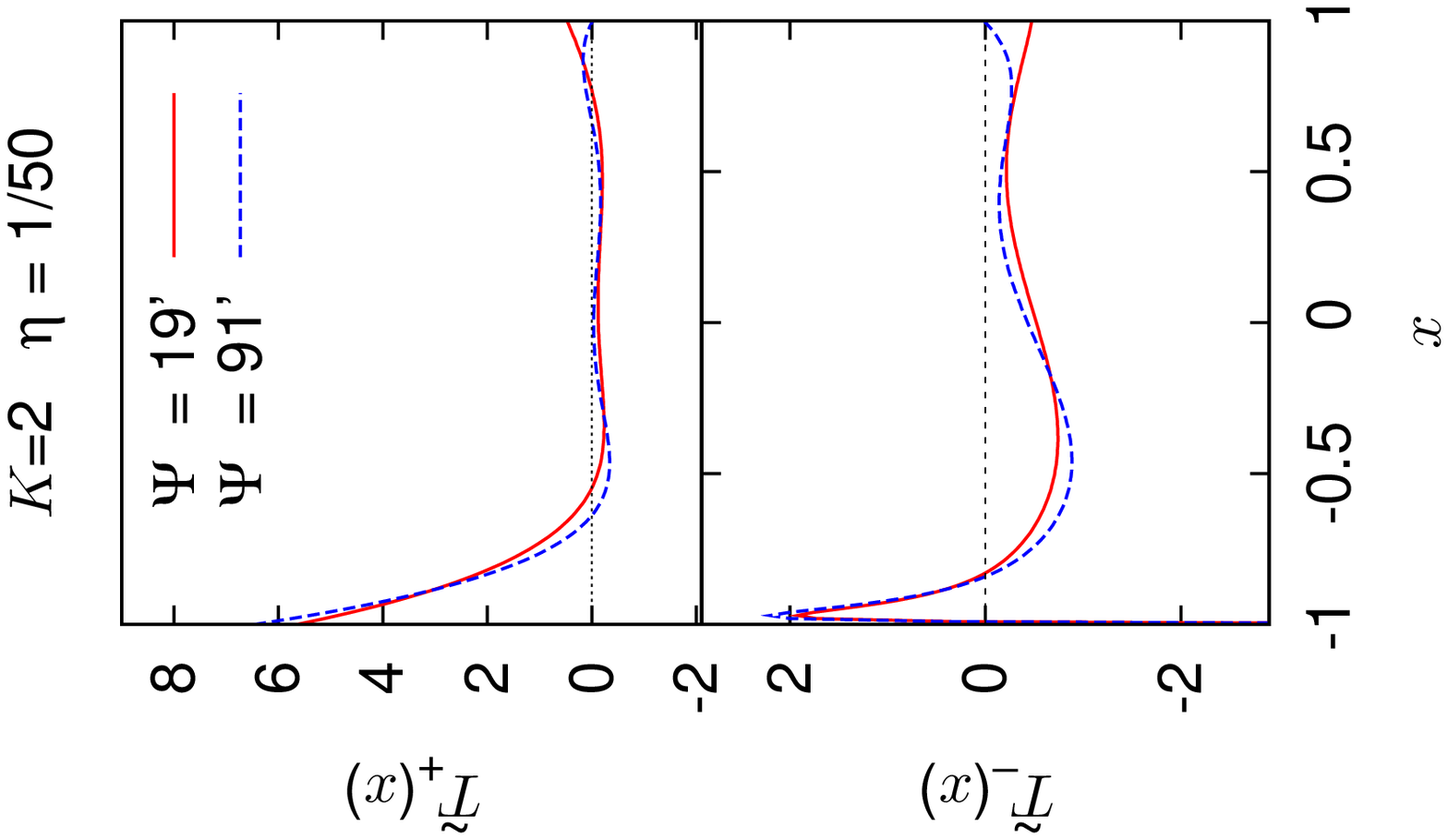}
}
    \end{center}

    \caption{The normalised functions $\tilde T_+$ and $\tilde T_-$
      optimised for signal-to-noise ratio at scales $\Psi = 19\arcmin$
      and $91\arcmin$, respectively.  \emph{Left two panels}: $\tilde
      T_+$ decomposed into polynomials of order $N=6$ and 15,
      respectively, for fixed $\vt_{\rm min}=0.2\arcmin$; \emph{Right
        two panels}: The comparison of fixed $\eta=1/10$ and $1/50$ in
      the case of polynomials of order $N=6$. }

  \label{fig:Tpm}
\end{figure*}

\section{Results}
\label{sec:results}

In this section we define $Q$ according to signal-to-noise and a
Fisher-matrix figure of merit, respectively, which we maximise to find
the corresponding optimised filter function $\tilde T_+$.  Before
that, we comment on our choice of $K$ and $N$, the number of
constraints and polynomial order, respectively.

\subsection{Number of constraints and polynomial order}
\label{sect:constraints}

We choose the minimum number of constraints $K=2$ necessary for a
finite-interval E- and B-mode decomposition, corresponding to the two
integral constraints (Eq.~\ref{c1c2}). Adding the two continuity
constraints (Eq.~\ref{Continuity}) resulted in significantly lower
values of $Q$. Some of the resulting functions $\tilde T_\pm$ showed
strong variations and narrow peaks for $|x|$ near unity. This
indicates that the continuity constraint is not very ``natural'' but
represents a strong restriction on the optimised filter functions. The
price that has to be paid for continuity is then a function which
fluctuates strongly, which may be problematic when applied to noisy
data.

Larger values of $N$ improved $Q$ to some extend but at the same time
increased the occurrence of local maxima found by the search
algorithm, reducing stability and reproducibility of the results. The
functions $\tilde T_\pm$ sometimes showed a high number of
oscillations.  A good choice for the polynomial order $N$ was found to
be 6, equivalent to 4 free parameters $a_n$ for $K=2$. A lower $N$
resulted in significantly smaller $Q$.

For the remainder of this paper we therefore choose $N=6$, $K=2$, if
not indicated otherwise.

\subsection{Shear covariance and cosmology}
\label{sec:model}

For the optimisation process we rely on a model shear correlation
function and covariance. The fiducial cosmology for our model is a
flat $\Lambda$CDM Universe with $\Omegam = 0.25, \Omegab = 0.044,
h=0.7$ and $\sigma_8 = 0.8$. We use the non-linear fitting formula of
\citet{2003MNRAS.341.1311S} together with the
\citet{1998ApJ...496..605E} transfer function for the matter power
spectrum. The redshift distribution of source galaxies is the best-fit
model of \citet{FSHK08} which has a mean redshift of 0.95.

We use the covariance matrix $C_{++}$ of the shear correlation
function $\xi_+$ from \cite{FSHK08}, corresponding to the third data
release of CFHTLS-Wide. This includes ellipticity noise and cosmic
variance as well as the residual B-mode added in quadrature. The
Gaussian part of the covariance was calculated using the method from
\citet{KS04}. Non-Gaussian corrections on small scale were applied
according to \citet{2007MNRAS.375L...6S}.

The $\RE$-covariance matrix, $\langle \RE^2 \rangle$, is an integral
over $C_{++}$,

\begin{equation}
  \left\langle \RE^2(\Psi_1, \Psi_2) \right\rangle =
  \int_{\vt_{{\rm min}, 1}}^{\Psi_1} \dd \vt \, \vt T_+(\vt)
  \int_{\vt_{{\rm min},2}}^{\Psi_2} \dd \vt' \, \vt' T_+(\vt') C_{++}(\vt,
  \vt').
      \label{RE_covar}
\end{equation}
We use the two upper scale limits $\Psi_1$ and $\Psi_2$ as arguments
of $\langle \RE^2 \rangle$; it also depends on the two lower scale
limits $\vt_{{\rm min},1}$ and $\vt_{{\rm min},2}$.

\subsection{Signal-to-noise ratio}

The first criterion for which we optimise the filter function is the
signal-to-noise ratio
\begin{equation}
  \rm {S/N}(\Psi) = \frac{\RE(\Psi)}{\left\langle
      \RE^2(\Psi,\Psi)\right\rangle^{1/2}}.
  \label{SN}
\end{equation}
The variance $\langle{\RE^2(\Psi,\Psi)}\rangle$ is the diagonal of
Eq.~(\ref{RE_covar}). As mentioned before, the signal-to-noise ratio
does not depend on the normalisation of $\tilde T_+$
(Eq.~\ref{normalisation}).

\subsubsection{Signal-to-noise for fixed $\vartheta_{\rm min}$}
\label{sec:SN_fixed_vt_min}

The signal-to-noise is calculated as a function of $\Psi = \vt_{\rm
  max}$, keeping $\vartheta_{\rm min}$ constant. We choose $\vt_{\rm
  min} = 0.2\arcmin$ which is a typical (albeit conservative) lower
limit where galaxies from ground-based data can be well separated. For
each scale $\Psi$ we obtain an optimised filter function $T_+$, as
discussed in Sect.~\ref{sec:optimisation} (see Table
\ref{tab:optimise}).

As can be seen in the left panel of Fig.~\ref{fig:SN_fixed_thetamin}
the optimal filter function with $N=6$ and $K=2$ results in a much
higher signal-to-noise than for original ring statistic function
$\RR$, using the filter function $Z_+$ from SK07. The new filter is
superior to the aperture-mass dispersion.  Note that we plot S/N for
$\langle M_{\rm ap}^2 \rangle$ as function of the aperture diameter
instead of the radius, to have the same maximum shear correlation
scale $\Psi$ as for $\RE$.

Although the optimal signal-to-noise is expected to be monotonic as
function of $\Psi$, this is clearly not the case.  However, when
increasing the polynomial order $N$ to 15, the signal-to-noise is
nearly constant for $\Psi > 19\arcmin$ and larger than for $N=6$ (see
Fig.~\ref{fig:SN_fixed_thetamin}). As a drawback, the S/N-curve for
$N=15$ is less smooth due to the difficulty of finding the global
maximum.  We did in general not find significantly larger values for
S/N with $N$ larger than 15.

The similar shape of S/N for the different cases $\RE$, $\langle
M_{\rm ap}^2 \rangle$ and $\RR$ is an imprint of the covariance
structure of the shear correlation function. The shape is modified
stronger for high $N$, where the peak at around 20 arc minutes
disappears.

The shear function $\RE$ is plotted in the right panel of
Fig.~\ref{fig:SN_fixed_thetamin}. It has a similar shape as $\RR$ from
SK07, and also as $\langle M_{\rm ap}^2 \rangle$. This is reflecting
the fact that all functions correspond to narrow filters and are
band-pass convolution of the power spectrum.

The normalised optimal filter functions for two angular scales are
shown in the left two panels of Fig.~\ref{fig:Tpm}.  The similarity of
the functions for different scales shows the relatively weak
dependence of the filter shape on angular scale.

\begin{figure*}

  \resizebox{\hsize}{!}{
    \includegraphics[angle=270,bb=50 100 554 760]{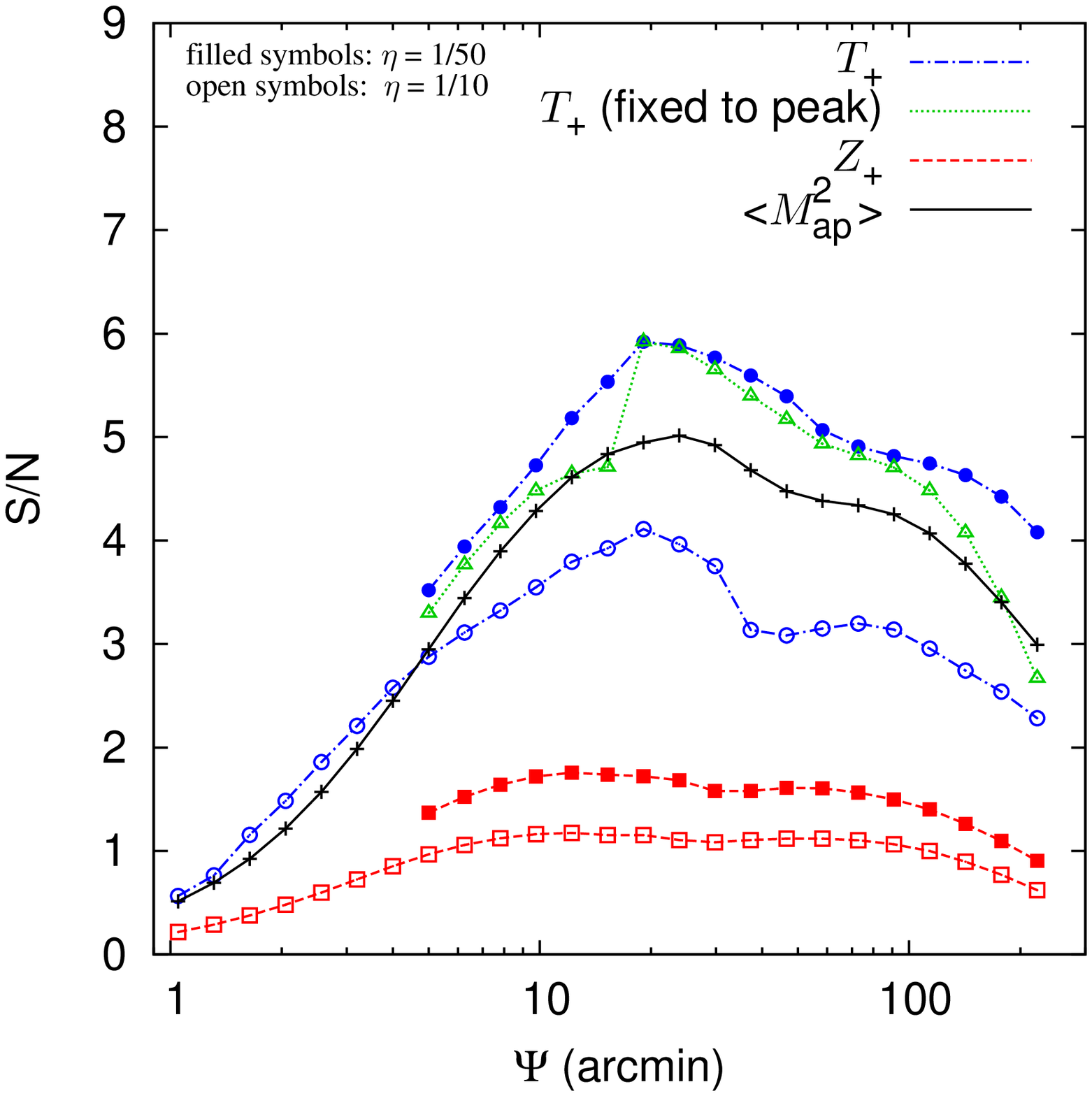}
    \includegraphics[angle=270,bb=50 115 554 770]{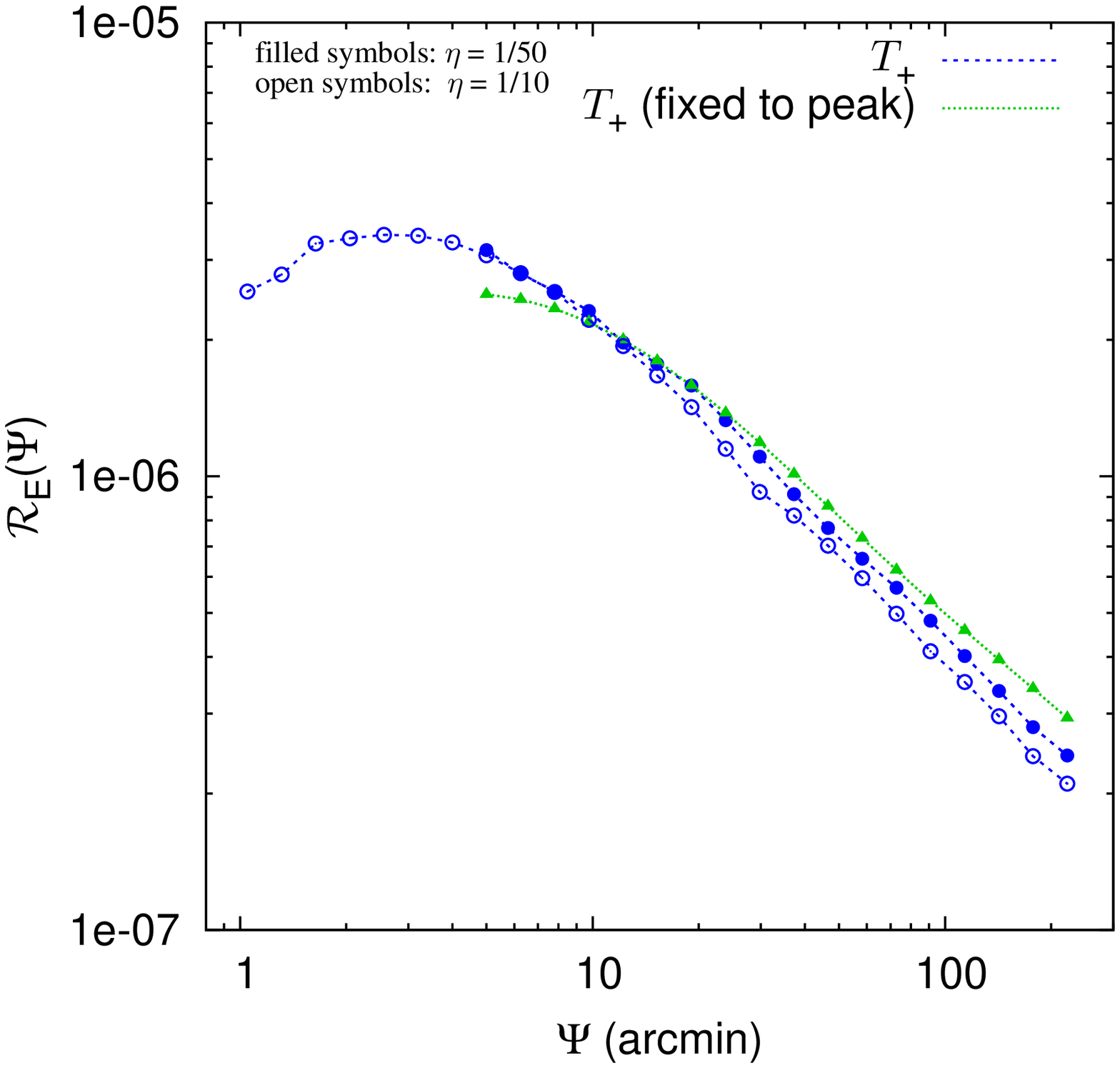}
  }
  \caption{\emph{Left panel}: Signal-to-noise ratio as function of
    $\Psi$, for fixed values of $\eta$ with $N=6$ and $K=2$. The blue
    curves correspond to the optimised function $\tilde T_+$. The
    green curve is obtained by applying the the optimal function $T_+$
    of $\Psi = 19\arcmin$ to all scales, instead of optimising each
    time separately, as done for the blue lines. The red curves shows
    the filter function $Z_+$ from SK07. The black curve with crosses
    corresponds to the aperture-mass with aperture diameter $\vt_{\rm
      max}$. Filled and open symbols represent $\eta = 1/50$ and
    $1/10$, respectively. \emph{Right panel}: $\RE$ for fixed values
    of $\eta$ of $1/10$ (open circles) and 1/50 (filled).  For the
    green line (with filled triangles), the optimal function $\tilde
    T_+$ for $\Psi = 19\arcmin$ was used on all scales, instead of
    optimising each time separately as for the blue lines.}

  \label{fig:SN_fixed_eta}
\end{figure*}

\begin{figure}

  \resizebox{\hsize}{!}{
    \includegraphics[angle=270,bb=50 90 554 720]{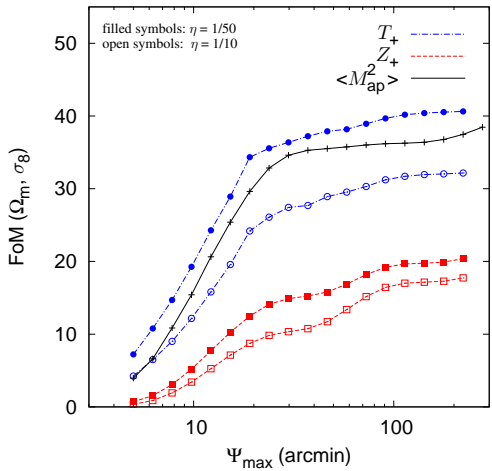}
  }

  \caption{Figure of merit (FoM, Eq.~\ref{FoM}) for the optimised
    filter function $T_+$ with $\eta=1/10$ and $\eta=1/50$ (blue
    curves). The red curves shows the FoM for the filter function $Z_+$
    from SK07. The black curve with crosses corresponds to the
    aperture-mass with aperture diameter $\vt_{\rm max}$. Filled and
    open symbols represent $\eta = 1/50$ and $1/10$, respectively.}
  \label{fig:fisher_fixed_eta_50}
\end{figure}

\begin{figure}

  \resizebox{\hsize}{!}{
    \includegraphics[angle=270,bb=80 60 480 750]{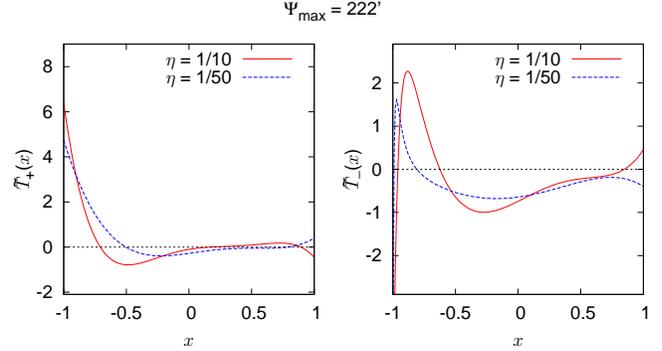}
  }

  \caption{The normalised functions $\tilde T_+$ and $\tilde T_-$,
    optimised for the Fisher matrix figure of merit (Eq.~\ref{FoM})
    with $\Psi_{\rm max} = 222\arcmin$. Two cases, $\eta=1/10$ and
    1/50 are shown.  }
  \label{fig:Tp_FoM}
\end{figure}

\subsubsection{Signal-to-noise for fixed $\eta$}
\label{sec:SN-fixed-eta}

Instead of fixing $\vt_{\rm min}$, we now leave $\eta$ constant and
change $\vt_{\rm min}$ along with $\Psi$. The signal-to-noise ratio
increases with decreasing $\eta$ (left panel of
Fig.~\ref{fig:SN_fixed_eta}). This is not surprising since a larger
$\eta$ means a smaller range of angular scales. For $\eta = 1/50$ the
optimal S/N exceeds the one using the aperture-mass dispersion.  The
optimal filter functions have similar shape to the previous case of a
fixed $\vartheta_{\rm min}$ (see the right two panels of
Fig.~\ref{fig:Tpm}).  We use the previous maximum coefficients $a_n$
as starting point for subsequent scales as discussed in
Sect.~\ref{sec:optimisation} (see Table \ref{tab:optimise}).  Unlike
in the previous case of fixed $\vt_{\rm min}$ we do not expect S/N to
be monotonous as function of $\Psi$, because $\vt_{\rm min}$ increases
with $\Psi$.

Since $\eta$ is constant, each filter function provides a valid
E-/B-mode decomposition for any given scale. We use the filter
optimised for $\Psi = 19\arcmin$, where the highest signal-to-noise
occurs, and apply it to the other scales (see the green curve with
triangles in the left panel of Fig.~\ref{fig:SN_fixed_eta}). As
expected, the signal-to-noise for scales $\Psi \ne 19\arcmin$ is lower
than in the previous case, where the optimisation was done for each
scale individually. The difference however is not large and this case
of fixed $\eta$ shows a S/N which is mostly larger than the
aperture-mass dispersion.

The shear function $\RE$ is plotted in the right panel of
Fig.~\ref{fig:SN_fixed_eta}, it is very similar in shape as in the
case of fixed $\vt_{\rm min}$ (Sect.~\ref{sec:SN_fixed_vt_min}).
Table \ref{tab:anlist} shows the polynomial coefficients of the
corresponding filter function $\tilde T_+$.

\subsection{Fisher matrix}
\label{sec:fisher}

\begin{figure*}

  \resizebox{\hsize}{!}{
    \begin{minipage}[c]{0.4\textwidth}
      \begin{center}Covariance matrix of $\RE$\end{center}
      \includegraphics[scale=0.6,bb=25 0 375 300]{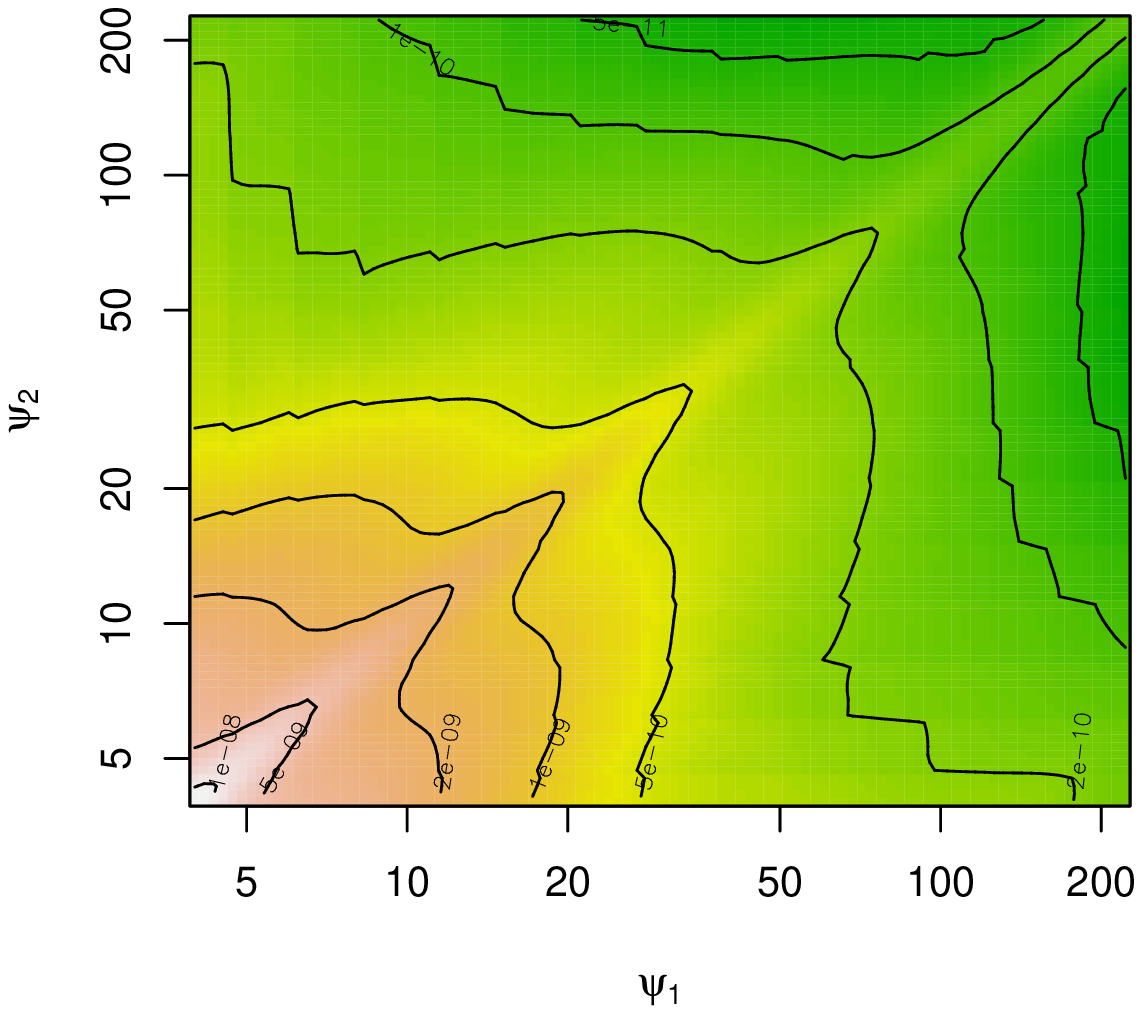}
    \end{minipage}
    \begin{minipage}[c]{0.4\textwidth}
      \begin{center}Correlation matrix of $\RE$\end{center}
      \includegraphics[scale=0.6,bb=25 0 375 300]{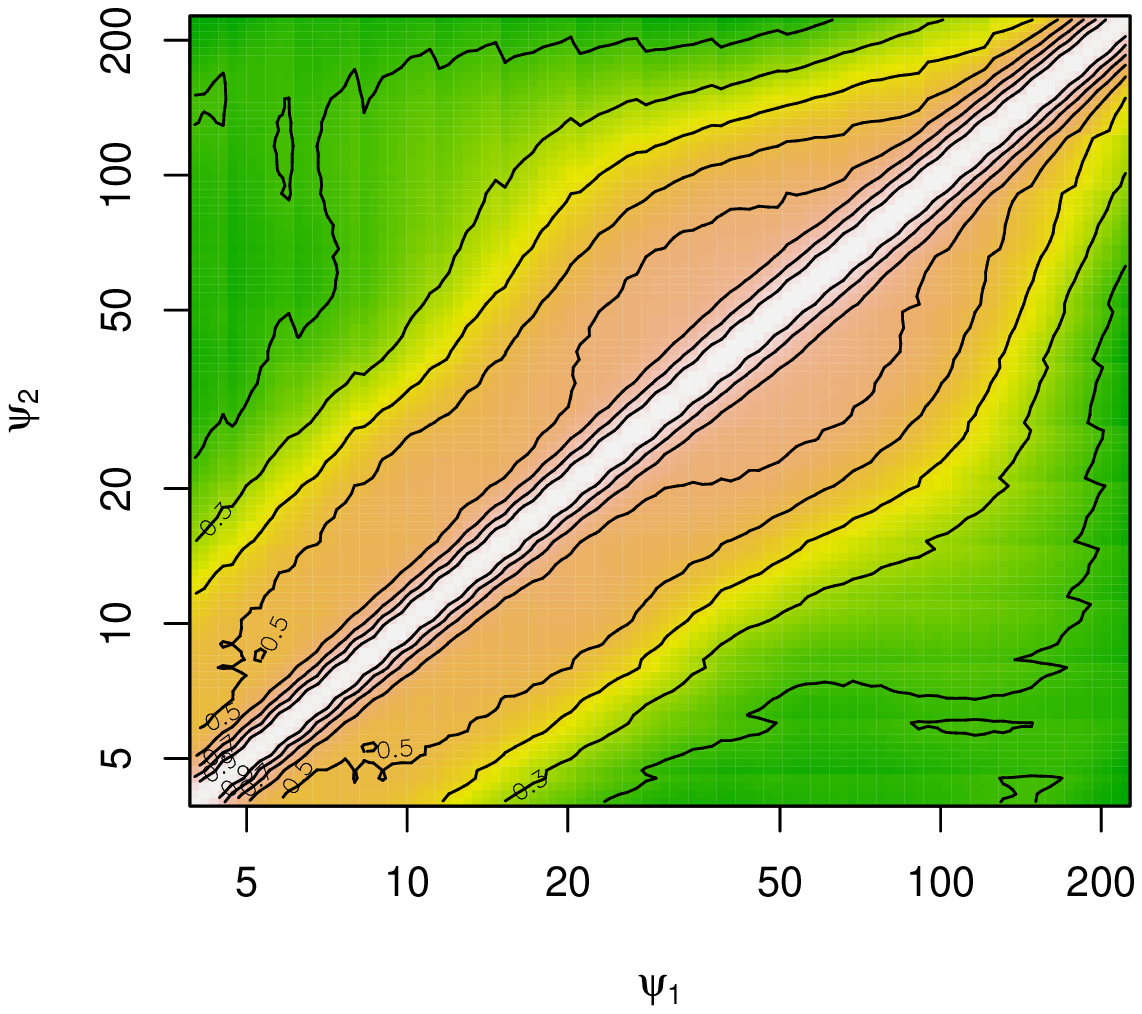}
    \end{minipage}
  }
  \resizebox{\hsize}{!}{
    \begin{minipage}[c]{0.4\textwidth}
      \begin{center}Covariance matrix of $\langle M_{\rm ap}^2 \rangle$\end{center}
      \includegraphics[scale=0.6,bb=25 0 375 300]{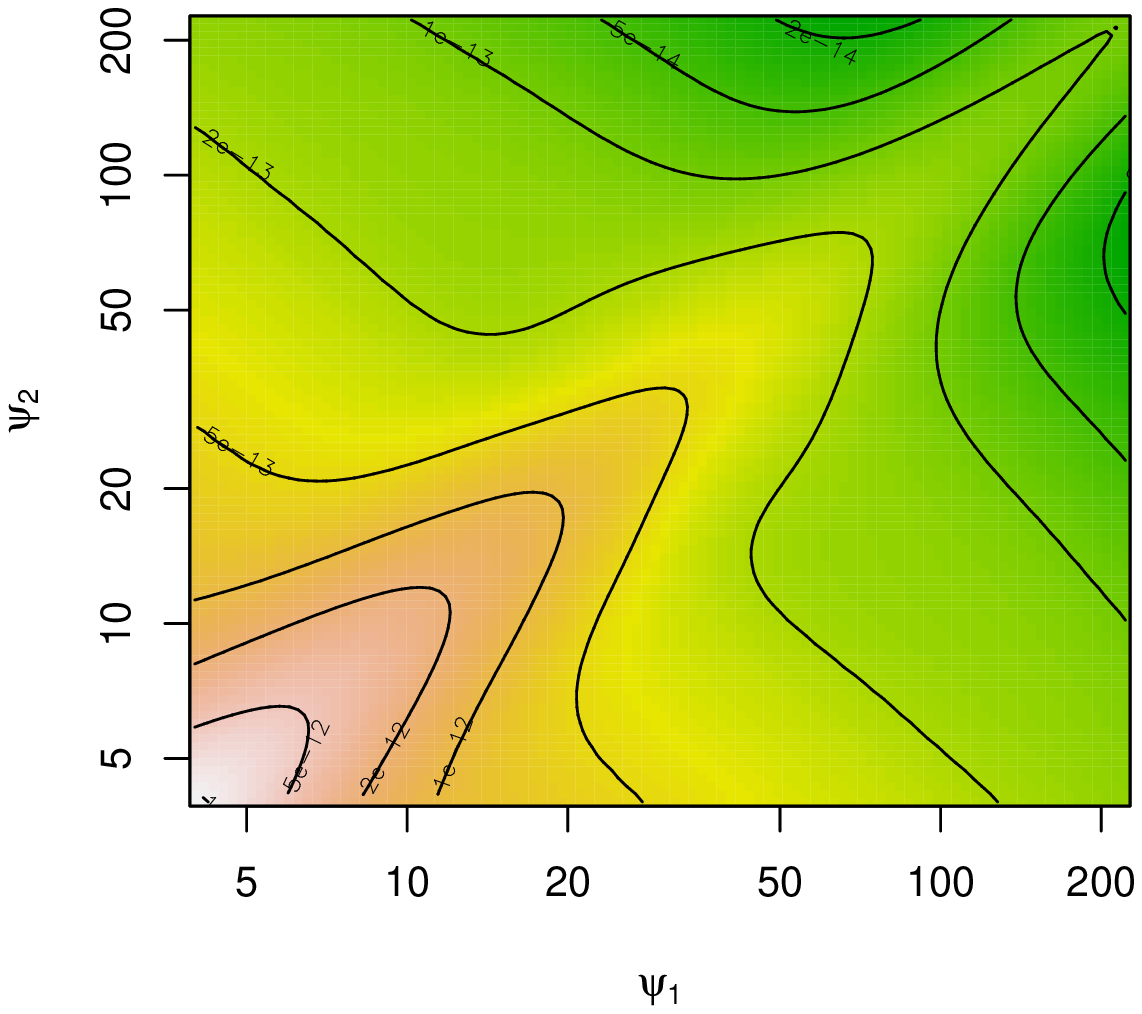}
    \end{minipage}
    \begin{minipage}[c]{0.4\textwidth}
      \begin{center}Correlation matrix of $\langle M_{\rm ap}^2 \rangle$\end{center}
      \includegraphics[scale=0.6,bb=25 0 375 300]{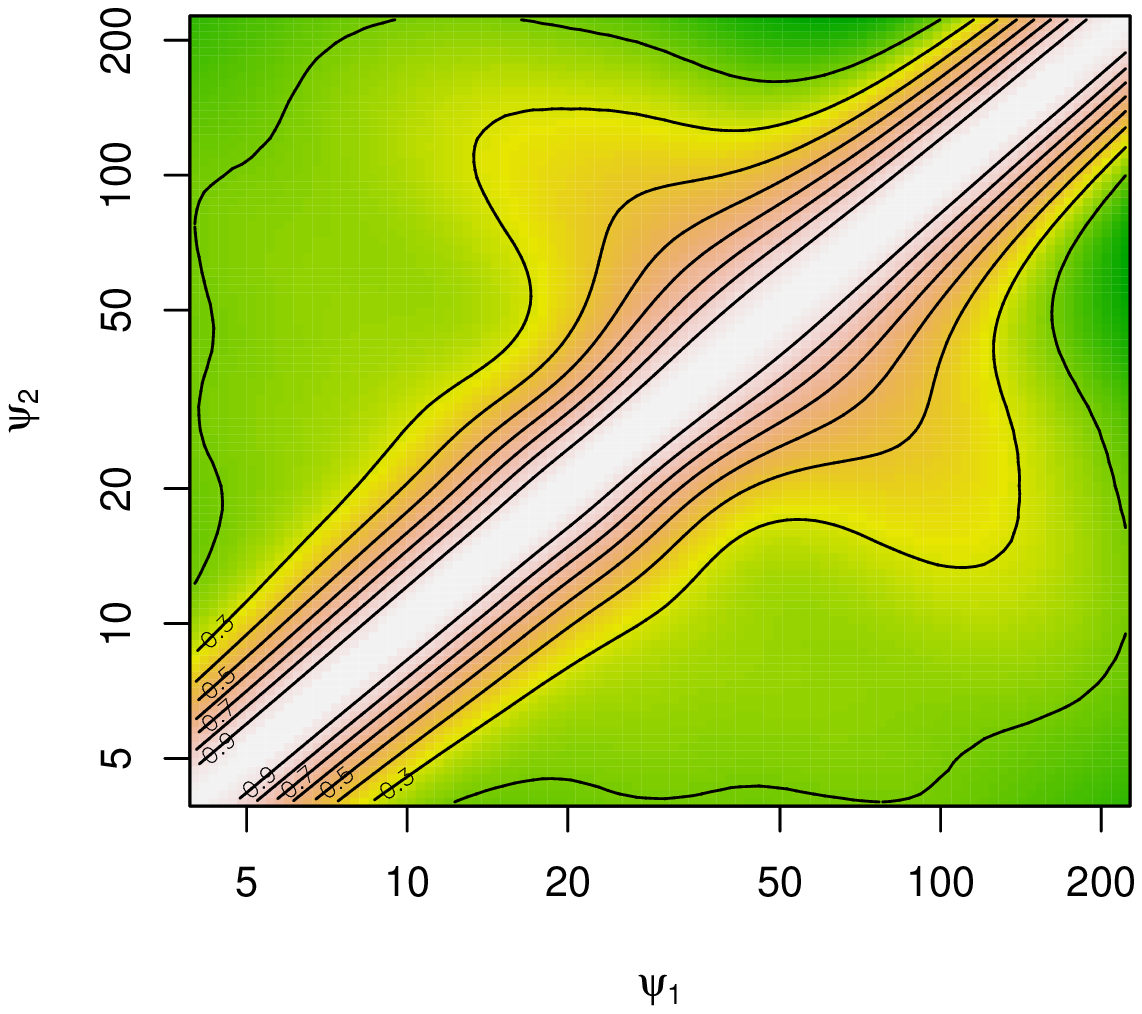}
    \end{minipage}
  }

  \caption{Covariance matrix (\emph{left panels}) and correlation
    matrix (\emph{right}) of $\RE$, optimised for the figure of merit
    (Eq.~\ref{FoM}) with $\eta=1/50$ (\emph{top}), and of the
    aperture-mass dispersion  with $\Psi_i$ being equal to the aperture
    diameter (\emph{bottom}). The colors correspond to
    the same levels for $\RE$ and $\langle M_{\rm ap}^2 \rangle$.  In
    the right panels, the contour lines start from the innermost value
    of 0.9 and are spaced by 0.1.  }
  \label{fig:cov_r_RR-fisher50}
\end{figure*}

To optimise the filter function we take now the alternative approach
of minimising the errors on cosmological parameters from our new
second-order shear statistic. To that end, we use the Fisher matrix
$F$, given by
\begin{equation}
  F_{\alpha\beta} = \sum_{ij}  \left\langle \RE^2(\Psi_i, \Psi_j)\right\rangle^{-1} \frac{\partial \RE(\Psi_i)}{\partial
    p_\alpha} \frac{\partial \RE(\Psi_j)}{\partial p_\beta}
 .
  \label{fisher}
\end{equation}
The cosmological parameters are comprised in the vector $\vec p$.  As
in the case of the signal-to-noise ratio, the Fisher matrix is
independent of the normalisation of the filter function $\tilde T_+$.

The quantity $Q$ to be maximised is the inverse area of the error
ellipsoid in parameter space, given by the Fisher matrix. In two
dimensions this figure of merit \cite[FoM,][]{DETF} is
\begin{equation}
  {\rm FoM}^{-1} = \pi \left( \sigma_{11} \sigma_{22} - \sigma_{12}^2
  \right)^{1/2}; \;\;\;\; \sigma_{ij}^2 = \left( F^{-1} \right)_{ij}.
  \label{FoM}
\end{equation}
\cite{ES09} used the quadrupole moment determinant $q$ of the
likelihood function to quantify the size of the parameter confidence
region. In case of a Gaussian likelihood (which correspond to our
Fisher matrix approximation) in two dimensions, the relation FoM$^{-1}
= \pi q$ holds.

We keep $\eta$ constant, allowing for a single filter function $\tilde
T_+$ to provide the E- and B-mode decomposition for each scale $\Psi$
in Eq.~(\ref{fisher}), and also for the covariance matrix. This
requirement is not a necessity since the covariance between scales can
be easily generalised to different filter functions. However, we
choose this approach for simplicity.  The cosmological parameters we
consider are $\Omegam$ and $\sigma_8$.

In Fig.~\ref{fig:fisher_fixed_eta_50} we compare the figure of merit
for the optimised filter function and the aperture-mass
dispersion. For a given maximum scale $\Psi_{\rm max}$ we vary
$\Psi_i$ in Eq.~(\ref{fisher}) between $4.0\arcmin$ and $\Psi_{\rm
  max}$. With $\eta = 1/50$, the minimum angular scale is
${\min}(\vt_{\rm min}) = 4.8$ arc second. For $\eta=1/10$ we use the
same angular scales, starting with 4.0\arcmin, for consistency with
the case $\eta=1/50$. Alternatively, by using the same minimum
angular scale of 4.8 arc second, the smallest $\Psi$ can be in
principle as small as $4.8\arcsec/\eta = 0.8\arcmin$. This addition
of small scales results in a higher FoM which is comparable to the
one for $\langle M_{\rm
  ap}^2 \rangle$.

We choose the same range of scales for the aperture-mass dispersion,
i.e.~we vary the aperture diameter between $4.0\arcmin$ and $\Psi_{\rm
  max}$. Note that although the minimum angular scale is theoretically
zero, in practise we are limited by the smallest scale for which the
$\xi_+$-covariance matrix is calculated which is $3\arcsec$ in our
case.

The normalised optimal filter functions $\tilde T_\pm$ are shown in
Fig.~\ref{fig:Tp_FoM}. They have a similar shape as the functions
optimised for S/N (see Fig.~\ref{fig:Tpm}). Table \ref{tab:anlist}
shows the corresponding polynomial coefficients.

\subsection{The covariance of $\RE$}
\label{sec:covariance}

\begin{figure}

  \resizebox{\hsize}{!}{
    \includegraphics{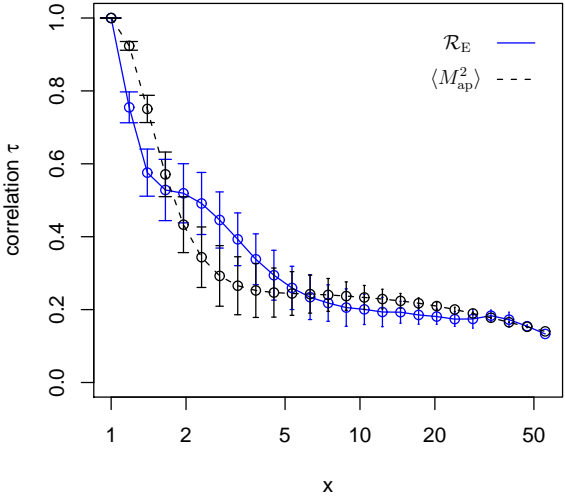}
  }

  \caption{The average correlation $\tau$ (Eq.~\ref{tau}) for scales which
    are separated by the ratio $x$, for $\RE$ (solid, blue lines) and
    $\langle M_{\rm ap}^2 \rangle$ (dashed, black curve). The error
    bars indicate the scatter when averaging over different scales $\Psi$.}
  \label{fig:len}
\end{figure}

We calculate the covariance matrix $C$ of $\RE$ using the optimal
filter function from the figure-of-merit maximisation at the largest
scale $\Psi_{\rm max} = 222\arcmin$, for a constant $\eta = 1/50$, see
Sect.~\ref{sec:fisher}. As can be seen in
Fig.~\ref{fig:cov_r_RR-fisher50} the covariance is
diagonally-dominated, similar to the one of the aperture-mass
dispersion. The degree of correlation is seen more clearly by
regarding the correlation matrix
\begin{equation}
  r(\Psi_1, \Psi_2) = \frac{\langle \RE^2(\Psi_1, \Psi_2)
    \rangle}{\langle \RE^2(\Psi_1, \Psi_1) \rangle \langle \RE^2(\Psi_2, \Psi_2) \rangle},
\end{equation}
see the right panels of Fig.~\ref{fig:cov_r_RR-fisher50}. To quantify
the correlation length, we compute the following function,
\begin{equation}
  \tau(x) = \left\langle r(\Psi, x \Psi) \right\rangle_\Psi,
  \label{tau}
\end{equation}
which is the correlation between two scales separated by the
multiplicative factor $x$, averaged over all $\Psi$. This function is
shown in Fig.~\ref{fig:len}. Since the lines of equal $r$ are mainly
parallel to the diagonal, the scatter is relatively small. The
correlation of $\RE$ drops off faster near the diagonal than the one
for $\langle M_{\rm ap}^2 \rangle$, and shows a slightly larger
correlation at intermediate distances $x$.  The covariance of $\RR$
has been studied in \citet{ES09} and has significantly smaller
correlation length than the one for $\langle M_{\rm ap}^2 \rangle$.

\subsection{Numerical limits on the B-mode $\RB$}
\label{sec:Bmode}

We calculate the B-mode $\RB$ from Eq.~(\ref{EB}) by using $\tilde
T_-$ obtained from the optimal function $\tilde T_+$ (see
Sect.~\ref{sec:Tminus}). Theoretically, $\RB$ vanishes but there could
be a residual B-mode because the E-/B-mode decomposition might not be
perfect. For example, there could be numerical issues regarding the
matrix inversion of Eq.~(\ref{eqK}). Our values of $\RB$ are limited
by the precision of the numerical integration of Eq.~(\ref{EB}). $\RB$
goes to zero for decreasing integration step size and we did not find
evidence for a residual B-mode. For a step size of $\Delta \vt = 5
\cdot 10^{-4}$ arc seconds, we find $\RB/\RE < 2\cdot 10^{-5}$ for all
angular scales. Even if a residual B-mode should be present, it is
straightforward to make it vanish identically. One can increase the
polynomial order $N$ of the decomposition by one, and determine the
corresponding coefficient $a_{N-1}$ such that $\RB = 0$.

\begin{table}
  \caption{Coefficients $a_n$ of the optimised function $\tilde T_+$ for (1)
    S/N, $\Psi=19\arcmin, \eta=1/50$; (2) FoM, $\Psi_{\rm max} =
    222\arcmin, \eta=1/10$; (3) FoM, $\Psi_{\rm max} =
    222\arcmin, \eta=1/50$.}
  \begin{tabular}{l|l|ll}
 \hline\hline
    & \rul  $\phm \rm {S/N}$  ($\Psi=19\arcmin$) & \multicolumn{2}{c}{FoM ($\Psi=222\arcmin$)}
    \\ 
      $n$ & $\phm a_n$ $(\eta=1/50)$ & $\phm a_n$ $(\eta=1/10)$ & $\phm a_n$ $(\eta=1/50)$ \\ \hline
    0 & $\phm0.1197730890$ & $\phm 0.009877788826$ & $\phm 0.1239456383$ \\
    1 &  $-0.3881211865$   & $\phm 0.1061397843$   & $-0.3881431858$ \\
    2 & $\phm0.5212557875$ & $-0.4300211814$   & $\phm 0.5579593467$ \\
    3 &  $-0.3440507036$   & $\phm 0.5451016406$    & $-0.3679282338$ \\
    4 & $\phm0.2761305382$ & $-0.3372272549$   & $\phm 0.1540941993$ \\
    5 & $-0.07286690971$   & $\phm 0.1716983151$   & $\phm 0.01293361618$ \\ \hline\hline
  \end{tabular}
  \label{tab:anlist}
\end{table}

\subsection{Dependence on cosmology and survey parameters}

The results presented in this paper have been obtained by using a
specific covariance matrix of $C_{++}$, namely the one used in
\cite{FSHK08}, and by choosing a specific cosmology. Here, we briefly
describe how our results change when we modify these parameters.

First, we illustrate the dependence on the covariance matrix.  Instead
of using the full covariance, as was done in the previous sections, we
repeat the S/N-analysis by taking the diagonal shot-noise component
only. This noise origins from the intrinsic galaxy ellipticity
dispersion. As expected, the S/N and FoM increase substantially,
mainly because of the missing cross-correlation between angular scales
in the shear correlation function. For all three cases, $\RE, \langle
M_{\rm ap}^2 \rangle$ and $\RR$, the S/N increases monotonously with
$\Psi$ beyond the maximum scale and does not show a peak at around 20
arc minutes.  The relative trend between the three cases stays the
same.

We change the fiducial cosmological model by increasing $\sigma_8$
from 0.8 to 0.9. This results in an increase of S/N of a factor
between 1.3 and 1.4, which is roughly the same for $\RE, \langle
M_{\rm ap}^2 \rangle$ and $\RR$. Increasing the mean redshift from
0.95 to 1.19 caused the S/N to be higher by 1.8 to 2, in the same way
for all three cases. We conclude that the relative difference is not
dependent on cosmology or the redshift distribution.

We repeat the calculation of S/N by choosing a fixed $\vt_{\rm min}$
which is different from our standard value of 0.2 arc minutes. The
improvement of $\RE$ over $\RR$ decreases for decreasing $\vt_{\rm
  min}$, by about $\Delta$(S/N) = 0.1 for $\Delta\vt_{\rm min} =
0.1\arcmin$, averaged over all scales $\Psi$.  This might be because
$\RR$ shows less cross-correlation between scales which leads to a
larger gain when additional scales are included.  The gain of $\RE$
with respect to $\langle M_{\rm ap}^2 \rangle$ increases when lowering
$\vt_{\rm min}$, as expected, since the inclusion of more small scales
boosts the S/N. On average, the difference is 0.05 for each 0.1 arc
minute which leads to an asymptotic value of
[S/N($\RE$)]/[S/N($\langle M_{\rm ap}^2 \rangle$)] = 1.31.

To check the stability of the results, we add an independent, uniform
random variable between $-p$ and $p$ to each of the highest $N-K$
coefficients $a_n$ after the optimum has been found.  For each
randomisation we fix $a_0, \ldots, a_{K-1}$ as described in
Sect.~\ref{sec:constraints} to assure E-/B-mode separation. The S/N
and FoM are very robust against changes in the coefficients.  For both
$p=0.01$ and 0.1, the changes in S/N and FoM are on the order $p$ and
less.

\section{Summary}
\label{sec:summary}


We have introduced a new second-order cosmic shear function which has
the ability to separate E- and B-modes on a finite interval of angular
scales. This function is a generalisation of the recently introduced
``ring statistic'' (SK07). Providing the second-order E-/B-mode shear
field correlations, general filter functions are calculated and
optimised for a specific goal. In this paper, we considered the
signal-to-noise ratio (S/N) as function of angular scale, and a figure
of merit (FoM) based on the Fisher matrix of the cosmological
parameters $\Omegam$ and $\sigma_8$ as optimisation criteria.

Our method to find the optimal filter function consists in the
following steps:

\begin{enumerate}

\item[1.] Choose the polynomial order $N$ and number of constraints
  $K\ge 2$ and define a quantity $Q$ to be maximised (in this work the
  signal-to-noise ratio S/N and the Fisher matrix figure-of-merit
  FoM).

\item[2.] Draw a random starting vector of coefficients $a_K, \ldots
  a_{N-1}$.

\item[3.] For $m=0 \ldots K-1$ calculate $s_m$ (Eq.~\ref{s_m}).

\item[4.] Invert the constraints matrix equation (\ref{eqK}) to get the
  first $K$ coefficients $a_0 \ldots a_{K-1}$.

\item[5.] Compute the filter function $T_+$ (Eqs.~\ref{theta2x}, \ref{decomp})

\item[6.] Calculate the shear function $\RE$ (Eq.~\ref{RE-noB})
    and $Q$ (in this work Eqs.~\ref{SN}, \ref{fisher}).

  \item[7.] Maximise $Q$. At each iteration of the maximisation
    process, repeat steps 3.-6.

\end{enumerate}

We were able to improve both S/N and FoM substantially with respect to
the SK07 ring statistic. Moreover, we obtained better results than for
the aperture-mass dispersion $\langle M_{\rm ap}^2 \rangle$, even
though the latter formally extends to zero lag and includes therefore
more small-scale power.

We have adapted and optimised our new second-order statistic $\RE$ to
a specific cosmology and survey parameters such as area and depth. We
used a smallest scale of 0.2\arcmin which corresponds to the smallest
separation for which galaxy images can easily be separated using
ground-based imaging data. The cosmic shear-correlation covariance
corresponds to the CFHTLS-Wide third data release used for weak
cosmological lensing \citep{FSHK08}.  The coefficients corresponding to
the optimal filter functions can be found in Table \ref{tab:anlist}. A
C-program which calculates the filter functions and the shear
statistic $\RE$ is freely
available\footnote{\texttt{http://www2.iap.fr/users/kilbinge/decomp\_eb/}}.

 Our specific results can be applied to
other requirements, although the results will not be
optimal. Alternatively, the optimisation method described here can
easily be applied to any given survey setting. For space-based
surveys, where the galaxy-blending confusion limit is smaller than for
ground-based observations, the advantage of $\RE$ over $\langle M_{\rm
  ap}^2 \rangle$ is more pronounced.

\section{Outlook}
\label{sec:outlook}

The new cosmic shear functions $\RE$ and $\RB$ can be applied in
various ways in the context of detecting systematics in cosmic shear
data and for constraining cosmological parameters.

The optimisation for signal-to-noise is useful for detecting a
potential B-mode in the data. In this case the E-/B-mode decomposition
serves merely as a diagnostic of the observations. A significant
B-mode can be a sign for residuals in the PSF correction or
non-perfect shape measurement. It might also hint to an
astrophysically generated B-mode signal, for example from shape-shear
correlations or shape-shape intrinsic alignment. In both cases, the
B-mode signal is expected to be small and it is of great importance to
obtain a clear E-/B-mode separation without mixing of modes.

In case of a suspected astrophysical B-mode, the separation of the
shear field into E- and B-modes might be a decisive advantage. If the
power spectrum contains both an E- and B-mode, $P_{\rm tot} = P_{\rm E}
+ P_{\rm B}$, both modes mix together into their Fourier-transform of
$P_{\rm tot}$, the shear correlation function. Thus, the E- and B-mode
may not be uniquely reconstructed from the correlation functions. The
different astrophysical components giving rise to $P_{\rm E}$ and
$P_{\rm B}$ can then only be separated by a E-/B-mode separating filter.

The figure-of-merit optimisation permits $\RE$ to be used for
efficient constraints on cosmological parameters. The reason to use a
filtered version of the shear correlation function $\xi_\pm$ instead
of the latter directly can be manifold \citep{EKS08}. A filter can be
chosen to be a narrow pass-band filter of the power spectrum and is
therefore able to probe its local features, unlike the broad low-pass
band function $\xi_+$. As a consequence, the correlation length is
much smaller and the covariance matrix close to diagonal.  This has
numerical advantages in particular in the case of many data points,
e.g.~for shear tomography. Finally, higher-order moments (skewness,
kurtosis, $\ldots$) of filtered quantities are easier to handle than
higher-order statistics of the (spin-2) shear field
\citep{JBJ04,SKL05}.

Apart from S/N and FoM maximisation, one can think of other,
alternative quantities with respect to which the filter function can
be optimised. For example, if a model for the B-mode is assumed, the
signal-to-noise of $\RB$ can be optimised to facilitate the possible
detection of a B-mode. Further, if cosmic shear is combined with other
probes of cosmology; the relative gain from weak lensing could be
maximised. This can be done for specific goals, for example a given
dark-energy parametrisation or some alternative theory of modified
gravity.  However, we emphasis that the possibilities are restricted
since the optimisation is always limited by the information contained
in the lensing power spectrum.

In the case of shear tomography, where the shear signal from different
redshifts is resolved (although only partially due to the broad
lensing efficiency kernel), one can perform a redshift-dependent
optimisation of the filter function. This is expected to bring further
improvements: Firstly, the projection of physical onto angular scales
varies with redshift; using a redshift-dependent filter function,
physical scales can be sampled optimally with redshift. Secondly, the
power spectrum changes with varying redshift; to optimise the sampling
of this redshift-dependent information might require a
redshift-varying filter.

$\RE$ is beneficial in particular on small scales, where the
aperture-mass dispersion suffers from mode-mixing. On scales less
than a few arc minutes there is a leakage of modes of about 10\%
\citep{KSE06}. Those scales contain information about halo
structure, substructure and baryonic physics. It is difficult to
model those effects, the use of those small scales to constrain
cosmological parameters is limited. On the other hand, lensing
observations on small scales will provide important constraints on
the physical processes involved and matter properties on small
scales.

\section*{Acknowledgments}

The authors want to thank Peter Schneider for fruitful discussions
and helpful comments during all stages of this work. We thank Ismael
Tereno, Mario Radovich, Yannick Mellier and Tim Eifler for useful
comments of the manuscript. LF acknowledge the support of the
European Commission Programme 6-th framework, Marie Curie Training
and Research Network ``DUEL'', contract number MRTN-CT-2006-036133.
MK is supported by the CNRS ANR ``ECOSSTAT'', contract number
ANR-05-BLAN-0283-04. MK thanks the Osservatorio di Capodimonte in
Naples for their hospitality. This project is partly supported by
the Chinese National Science Foundation Nos. 10878003 \& 10778725,
973 Program No. 2007CB 815402, Shanghai Science Foundations and
Leading Academic Discipline Project of Shanghai Normal University
(DZL805).

\appendix
\section{Simultaneous optimisation for arbitrary scales}
\label{sec:app}

The optimisation scheme introduced in this paper holds for a given
ratio of minimum and maximum scale $\eta = \vt_{\rm min}/\vt_{\rm
  max}$. In this section we introduce a simple generalisation of the
scheme to obtain an optimised function $\tilde T_+$ which fulfills the
integral constraints (Eq.~\ref{c1c2}) for all  $(\vt_{\rm min}, \vt_{\rm
  max})$. This comes at the expense of a poor resulting
signal-to-noise.

If we demand the following relation to hold
\begin{equation}
  I_\nu \equiv \int_{-1}^{+1} \dd x \, x^\nu \tilde T_+(x) = 0; \;\;\;\;
  \mbox{for} \;\;\;\; \nu=0,1,2,3,
  \label{rnu}
\end{equation}
then the two integral constraints
\begin{align}
  R I_0 + I_1 = R^3 I_0 + R^2 I_1 + R I_2 + I_3 = 0
\end{align}
 are satisfied. However, instead of two conditions we have now four
 equations (Eq.~\ref{rnu}) which fix $K=4$ coefficients of the
 decomposition. In this case, the first four matrix elements are
 (cf.~Eq.~\ref{faeq0})
\begin{equation}
  f_{mn} = \int_{-1}^{+1} \dd x \, x^m \, C_n(x); \;\;\;\; m=0 \ldots 3.
\end{equation}

Since there are two more integrals than in the previous case, the
resulting function has at least two more zeros. The corresponding
signal-to-noise ratio is significantly lower than in the
single-scale case; it is even smaller than the one obtained for
$Z_+$. We therefore do not consider this option further.

\bsp

\label{lastpage}


\end{document}